\documentclass[authoryear,preprint,review,12pt]{article}

\usepackage[a4paper, portrait, margin=1.1811in]{geometry}
\usepackage[english]{babel}
\usepackage[utf8]{inputenc}
\usepackage[T1]{fontenc}
\usepackage{helvet}
\usepackage[retainorgcmds]{IEEEtrantools}
\usepackage{etoolbox}
\usepackage{graphicx}
\usepackage{titlesec}
\usepackage{caption}
\usepackage{csquotes}
\usepackage{tabularx} 
\usepackage{booktabs}
\usepackage{xcolor} 
\usepackage[most]{tcolorbox}
\usepackage{tikz}
\usetikzlibrary{3d}
\usetikzlibrary{external}   
\usetikzlibrary{calc, decorations.pathreplacing, arrows.meta}

\usepackage{pgfplots}
\usepgfplotslibrary{}
\pgfplotsset{compat=newest}
\usepgfplotslibrary{fillbetween}

\usepackage{flafter} 
\usepackage[obeyFinal]{todonotes}
\usepackage{subcaption}
\usepackage{stmaryrd}
\usepackage{bm}
\usepackage{amsmath,amssymb, amsthm,mathtools,mathrsfs,stmaryrd,titletoc}
\usepackage{wasysym}
\usepackage[norefs,nocites,ignoreunlbld]{refcheck} 
\usepackage[acronym]{glossaries}
\usepackage{xfrac} 
\usepackage{nicefrac} 
\usepackage{numprint} 
\usepackage{physics}
\usepackage{listings} 
\usepackage{algorithm} 
\usepackage{algpseudocode} 
\usepackage{siunitx} 
\usepackage{threeparttable}
\usepackage{xspace}

\newcommand{\eg}{\textit{e.g.}~}

\newcommand{\Davila}{D\'{a}vila}

\newcommand{\jump}[1]{\ensuremath{\llbracket#1\rrbracket}}
\newcommand{\mc}[1]{\ensuremath{\left<#1\right>}}
\newcommand{\Dam}{\mathcal{D}}
\newcommand{\B}{\mathcal{B}}
\newcommand{\fatprops}{\mathcal{P}}
\newcommand{\hpdv}[2]{\ensuremath{\partial#1/\partial#2}}

\newcommand{\hfrac}[2]{\ensuremath{#1/#2}}

\newcommand{\dispjump}{\ensuremath{\llbracket\textbf{u}\rrbracket}}

\newcommand{\Npermm}[1]{\newton\per\milli\meter}
 
\usepackage[colorlinks, citecolor=cyan]{hyperref}
\hypersetup{
	colorlinks=true,
	linkcolor=blue,
	filecolor=magenta,      
	urlcolor=blue,
	citecolor=blue 
}

\usepackage[nameinlink]{cleveref}

 \AtBeginDocument{%
    \crefname{equation}{equation}{equations}%
    \crefname{chapter}{chapter}{chapters}%
    \crefname{section}{section}{sections}%
    \crefname{appendix}{appendix}{appendices}%
    \crefname{enumi}{item}{items}%
    \crefname{footnote}{footnote}{footnotes}%
    \crefname{figure}{figure}{figures}%
    \crefname{table}{table}{tables}%
    \crefname{theorem}{theorem}{theorems}%
    \crefname{lemma}{lemma}{lemmas}%
    \crefname{corollary}{corollary}{corollaries}%
    \crefname{proposition}{proposition}{propositions}%
    \crefname{definition}{definition}{definitions}%
    \crefname{result}{result}{results}%
    \crefname{example}{example}{examples}%
    \crefname{remark}{remark}{remarks}%
    \crefname{note}{note}{notes}%
}

\usepackage{caption}
\graphicspath{ {./images/} }
\graphicspath{{inkscape/}}
\usepackage{fancyhdr}
\usepackage{graphicx}

\fancypagestyle{plain}{
	\fancyhf{}

	\lhead{\color{cyan}\small \textbf{~}\\ \color{black}
	\textit{~}\\ }	
}

\usepackage[style=numeric,citestyle=phys,maxbibnames=9,maxcitenames=2,backend=biber, natbib=true, giveninits=true]{biblatex}

\patchcmd{\@maketitle}{\LARGE \@title}{\fontsize{16}{19.2}\selectfont\@title}{}{}
\makeatother

\usepackage{authblk}

\newsavebox\affbox
\author[]{\textbf{P. Hofman$^*$}}
\author[]{\textbf{F. P. van der Meer}}
\author[]{\textbf{L.J. Sluys}}
\affil[]{Delft University of Technology, Faculty of Civil Engineering and Geosciences, Delft, Netherlands}
\date{September 20, 2023}

\titlespacing\section{0pt}{12pt plus 4pt minus 2pt}{0pt plus 2pt minus 2pt}
\titlespacing\subsection{12pt}{12pt plus 4pt minus 2pt}{0pt plus 2pt minus 2pt}
\titlespacing\subsubsection{12pt}{12pt plus 4pt minus 2pt}{0pt plus 2pt minus 2pt}

\titleformat{\section}{\normalfont\fontsize{10}{15}\bfseries}{\thesection.}{1em}{}
\titleformat{\subsection}{\normalfont\fontsize{10}{15}\bfseries}{\thesubsection.}{1em}{}
\titleformat{\subsubsection}{\normalfont\fontsize{10}{15}\bfseries}{\thesubsubsection.}{1em}{}

\titleformat{\author}{\normalfont\fontsize{10}{15}\bfseries}{\thesection}{1em}{}

\title{\textbf{\Large A numerical framework for simulating progressive failure in composite laminates under high-cycle fatigue loading}\\}

\begin{document}

\pagestyle{headings}	
\newpage
\setcounter{page}{1}
\renewcommand{\thepage}{\arabic{page}}
	
\captionsetup[figure]{labelfont={bf},labelformat={default},labelsep=period,name={Figure }}	\captionsetup[table]{labelfont={bf},labelformat={default},labelsep=period,name={Table }}
\setlength{\parskip}{0.5em}
	
\maketitle
	
\noindent\rule{15cm}{0.5pt}
	\begin{abstract}
  In this work, a recently proposed high-cycle fatigue cohesive zone model, which covers crack initiation and propagation with limited input parameters, is embedded in a robust and efficient numerical framework for simulating progressive failure in composite laminates under fatigue loading.
  The fatigue cohesive zone model is enhanced with an implicit time integration scheme of the fatigue damage variable which allows for larger cycle increments and more efficient analyses. 
  The method is combined with an adaptive strategy for determining the cycle increment based on global convergence rates. 
  Moreover, a consistent material tangent stiffness matrix has been derived by fully linearizing the underlying mixed-mode quasi-static model and the fatigue damage update. 
  The enhanced fatigue cohesive zone model is used to describe matrix cracking and delamination in laminates. 
  In order to allow for matrix cracks to initiate at arbitrary locations and to avoid complex and costly mesh generation, the phantom node version of the eXtended finite element method (XFEM) is employed. 
  For the insertion of new crack segments, an XFEM fatigue crack insertion criterion is presented, which is consistent with the fatigue cohesive zone formulation.
  It is shown with numerical examples that the improved fatigue damage update enhances the accuracy, efficiency and robustness of the numerical simulations significantly. 
  The numerical framework is applied to the simulation of progressive fatigue failure in an open-hole [$\pm45$]-laminate. 
  It is demonstrated that the numerical model is capable of accurately and efficiently simulating the complete failure process from distributed damage to localized failure.

		\let\thefootnote\relax\footnotetext{
			\small $^{*}$\textbf{Corresponding author.} \textit{
				\textit{E-mail address: P.Hofman@tudelft.nl}\\
      }
    }
		\textbf{\textit{Keywords}}: \textit{composite laminates; high-cycle fatigue, progressive failure; extended finite element method, cohesive zone modeling}
	\end{abstract}

\section{Introduction}
Fatigue is often a critical failure process in fiber reinforced polymer laminates. Accurate, efficient and robust numerical prediction tools can help in enhancing the efficiency of the design process of composite laminated structures, reducing manufacturing time, and minimizing the need for extensive testing to ensure the safety of new composite structures. 

When fiber reinforced polymer laminates are subjected to loads, several interacting failure processes take place that make developing reliable prediction tools a challenging endeavor. For example, matrix cracks can initiate and propagate and eventually lead to interface delamination. Finally, fiber breakage can occur leading to overall failure of the laminate. The interaction of these competing processes and the type of final failure depend on the stacking sequence, laminate thickness and the presence of notches \cite{Wisnom2009, Green2007}. Furthermore, the load character (cyclic or quasi-static) influences the final failure mode \cite{Nixon-Pearson2013a,Nixon-Pearson2013a,Aymerich2000, Aidi2015}. In addition, damage accumulation in fiber reinforced composite materials already takes place at an early stage which results in significant stiffness reduction and stress redistribution. In order to accurately predict the performance of composite laminates, numerical models must take into account the progressive character of failure that covers the initial stage of early damage up to complete failure of the laminate. 

In literature, several numerical models have been developed for the quasi-static load case of complex laminates. For example, Jiang et al.~\cite{Jiang2007} modelled splitting and matrix cracks with interface elements at pre-defined locations according to experiments. Hallet et al.~\cite{Hallett2009} used a similar approach for matrix cracks and delamination and included a Weibull statistical criterion for fiber failure.  Furthermore, Van der Meer et al.~\cite{Meer2009, Meer2010} used the \textit{phantom node version of the extended finite element method} (XFEM) \cite{Hansbo2004} for modeling matrix cracks to reduce the complexity of meshing and to allow for mesh-independent cracks that can initiate at arbitrary locations with a pre-defined crack spacing. Moreover, \citeauthor{Chen2016} \cite{Chen2016} used a 3D version of the \textit{floating node method}~\cite{Chen2014} to model the interaction of a large number of discrete matrix cracks with delaminations and demonstrated that the model was able to accurately predict the sequence of failure processes in notched and unnotched laminates. More modeling approaches of composite laminates subjected to quasi-static loading scenarios can be found in Refs.~\cite{Swindeman2013, Chen2013, Achard2014, Kawashita2012a,Iarve2016a,Ma2021ProgressiveSimulation, Falco2018ModellingFramework, Le2018a, Chen2014SimulatingEffects, Wang2021}

At present, there are only few progressive failure models that can simulate (high-cycle) fatigue failure in composite laminates. One of the first papers of open-hole fatigue modeling of progressive failure was presented by Nixon-Pearson et al.~\cite{Nixon-Pearson2013b}, where pre-inserted interface elements were used to model matrix cracks and a Paris-type cyclic cohesive zone model (CZM) \cite{Harper2010c,Kawashita2012} was used for crack propagation. Iarve et al.~\cite{Iarve2016a} developed a model where the regularized eXtended finite element method (Rx-FEM) was employed for modeling mesh-independent cracks and a fatigue initiation criterion based on S-N curves was presented. This framework was further extended by \citeauthor{luFatigue2022}~\cite{luFatigue2022} to account for high-density crack networks by allowing cracks to appear close to each other. In the work by Tao et al.~\cite{Tao2018}, a similar approach as in \citeauthor{Nixon-Pearson2013b}~\cite{Nixon-Pearson2013b} was adopted, but extended with a fatigue initiation criterion based on S-N curves, similar as proposed in Ref. \cite{Iarve2016a}. As opposed to the discrete crack modeling approaches for simulating matrix cracking in the previously mentioned approaches, Llobet et al.~\cite{Llobet2021a} used a continuum damage modeling approach with fiber-aligned meshes and included a description for fiber damage due to cyclic loading. More recently, \citeauthor{Tao2023} \cite{Tao2023} developed an enhanced fatigue cohesive zone model where stiffness degradation is described by a Paris-relation-informed neural network and applied it to simulate an open-hole fatigue tension test with good accuracy. 

In most of the existing methods, a fatigue cohesive zone model, that requires a Paris-relation \cite{Paris1961} as input, is used for modeling delamination and matrix cracking. These cohesive zone models are suitable for describing crack propagation of an initial crack \cite{Bak2016, Harper2010c,Kawashita2012,Latifi2015a,Turon2007b,Carreras2019a,Amiri-Rad2017a,Amiri-Rad2017}. However, in full-laminate analysis, a matrix crack can also initiate under cyclic loading. Upon further applying fatigue load cycles, a fracture process zone develops in the onset phase after which propagation of the crack takes place. In literature, only a few CZMs take these three stages of fatigue crack growth into account \cite{Davila2020,Iarve2016a,luFatigue2022, May2010a,May2011,Nojavan2016,Llobet2021a}.

Recently, \Davila~\cite{Davila2020} proposed a cyclic CZM that covers initiation, onset and propagation and is built on Turon's quasi-static mixed-mode CZM~\cite{Turon2006ALoading, Turon2018b}. The fatigue CZM relies on S-N curves with simple engineering assumptions and empirical relations to take the dependence of mode-mixity and stress-ratio into account. The method is based on the assumption that an intrinsic relation exists between S-N curves and Paris' relation \cite{Allegri2020b}. The model is capable of simulating complex 3D crack fronts \cite{Raimondo2022} in a reinforced double cantilever beam (DCB) test \cite{carrerasBenchmarkTestValidating2019} and can be extended to cases where the \emph{local} stress ratio is not equal to the \emph{global} load ratio, as presented in \cite{Joosten2022} where the presence of residual stresses was taken into account. Furthermore, it has been demonstrated that the method can simulate crack migration in a ply-drop specimen \cite{liangReducedinput2021} and R-curve effects in thermoplastic material systems \cite{lecinanaCharacterization2023}. With regards to laminate analyses, the unification of initiation and propagation makes the fatigue CZM suitable for simulating \emph{both} interface delamination and matrix cracking in progressive failure analyses of complex laminates. 

Originally, the model presented in Ref.~\cite{Davila2020} employed an explicit update of the damage variable that evolves with load cycles. Therefore, small cycle increments must be used during simulations to prevent instabilities in the damage evolution and a step size criterion based on the maximum damage experienced in all integration points is required. This approach may become problematic in full-laminate analyses where many integration points, due to stress redistribution and complex loading histories, experience different damage rates throughout the simulation. 

In this work, an implicit fatigue damage update is presented to make the formulation more suitable for use in full-laminate analyses. The method is combined with efficient cycle jumping based on \emph{global} iterations and a fully consistent tangent matrix has been derived to enhance efficiency and robustness of the simulations. In order to allow for multiple cracks at arbitrary locations, the fatigue CZM is combined with XFEM with a proper fatigue crack insertion criterion that is consistent with the fatigue damage evolution.

The organization of this article is as follows. Firstly the improvement to D\'{a}vila's fatigue damage formulation is presented and a fully consistent material tangent is derived. Subsequently, the XFEM implementation with a proper fatigue crack insertion criterion is presented. The model is applied to several numerical examples to verify the methods and to the show the improved performance. In order to demonstrate the capabilities of the presented numerical framework, an open-hole [$\pm$45]-laminate under fatigue loading is simulated. 

\section{Methods}

\subsection{Fatigue cohesive zone model}
\label{sec:fczm-formulation}

The fatigue CZM by D\'{a}vila~\cite{Davila2020,Davila2020Nasa} is built on top of the static CZM with mode-dependent dummy stiffness by Turon \cite{Turon2006ALoading, Turon2018b}. In this section, the formulation of the fatigue CZM is given in local coordinate frame with basis vectors $\{\bm{e}_{n},\bm{e}_{s1},\bm{e}_{s2}\}$ aligned with the crack plane.

In order to allow for a reduction of elastic stiffness due to fatigue and static loading, a scalar damage variable $d$ is introduced. The traction is computed as

\begin{equation}
  \textbf{t} = (\textbf{I}-d\textbf{P})\textbf{K} \jump{\textbf{u}}
  \label{eq:traction-update}
\end{equation}
where $\dispjump$ is the displacement jump, $\textbf{K}$ is the dummy stiffness matrix and $\textbf{P}$ is a selection matrix expressed as

\begin{equation}
  \textbf{K} = \begin{bmatrix} 
    K_n  & 0 & 0 \\
    0 & K_{sh} & 0 \\
    0 & 0 & K_{sh} 
  \end{bmatrix},
  \label{eq:stiffness-tensor}
\end{equation}

\begin{equation}
  \textbf{P} = \begin{bmatrix} 
    \frac{\mc{\jump{u}_n}}{\jump{u}_n} & 0 & 0 \\
    0 & 1 & 0 \\
    0 & 0 & 1 
  \end{bmatrix} 
  \label{eq:selection-tensor}
\end{equation}
where $K_n$ and $K_{sh}$ are the normal and shear dummy stiffnesses respectively. The operator $\mc{\bm{\bullet}}$ is the Macaulay operator, defined as $\mathrm{max}(0,\bullet)$ and makes sure that interfacial penetration is prevented when the normal component of the displacement jump is negative. The damage variable $d$ determines the stiffness reduction and its evolution depends on the mode-mixity such that the energy dissipated matches with the phenomenological mixed-mode fracture energy relation proposed by Benzeggagh and Kenane \cite{Benzeggagh1996}. It is shown in \cite{Turon2010,Turon2018b} that the correct energy dissipation under mixed-mode fracture is ensured by relating the ratio between dummy stiffnesses $K_n$ and $K_{sh}$ to the fracture properties with the following equation 
\begin{equation}
  K_{sh} = K_n \frac{G_{Ic,d}}{G_{IIc,d}}\Bigg(\frac{f_{sh}}{f_n}\Bigg)^2
  \label{eq:constrain-dummies}
\end{equation}
where $f_{n}$, $f_{sh}$, $G_{Ic,d}$ and $G_{IIc,d}$ are the tensile strength, shear strength, mode-I and mode-II fracture energies, respectively. 
The mixed-mode CZM is formulated in the form of an equivalent 1D traction-separation relation

\begin{equation}
  \sigma = (1-d)K_\B \Delta
  \label{eq:1d-tsl}
\end{equation}
where $\sigma$ is the equivalent stress, $K_\B$ is the mode-dependent dummy stiffness and $\Delta$ is the equivalent displacement jump. These quantities are defined as
\begin{align}
  &\sigma = \sqrt{\mc{t_n}^2+ t^2_{s1} + t^2_{s2}} \label{eq:sigma-equiv}\\
  &K_{\B} = K_n (1-\B) + \B K_{sh}\label{eq:mode-dependent-dummy} \\
  &\Delta = \frac{K_n \mc{\jump{u}_n}^2 + K_{sh} \jump{u}_{sh}^2}{\sqrt{K_{n}^2\mc{\jump{u}_n}^2 + K_{sh}^2 \jump{u}_{sh}^2 }} 
\end{align}
where $\B$ is a displacement-based measure of mode-mixity

\begin{equation}
  \B = \frac{K_{sh} \jump{u}_{sh}^2}{K_{n} \mc{\jump{u}_n}^2 + K_{sh} \jump{u}_{sh}^2 } 
  \label{eq:mode-mixity-def}
\end{equation}
and $\jump{u}_{sh}$ is the Eucledian norm (length) of the shear displacement jump vector 
\begin{equation}
  \jump{u}_{sh}^2 = \jump{u}_{s1}^2 + \jump{u}_{s2}^2 
\end{equation}

The values of equivalent displacement jump at fracture initiation and complete fracture of the 1D equivalent traction-separation relation (\cref{eq:1d-tsl}) are expressed as
\begin{align}
  &\Delta_0 = \sqrt{\frac{K_n(\jump{u}^0_n)^2 + \Big(K_{sh}(\jump{u}^{0}_{sh})^2 - K_n(\jump{u}^0_{n})^2 \Big) \B^\eta}{K_{\B}}}  \label{eq:initiation-jump}\\
  &\Delta_f = \frac{K_n \jump{u}^0_n \jump{u}^f_n + \Big(K_{sh}\jump{u}^{0}_{sh} \jump{u}^{f}_{sh} - K_n \jump{u}^{0}_{n} \jump{u}^{f}_{n} \Big) \B^\eta}{K_{\B} \Delta_0} 
  \label{eq:equiv-jump-defs}
\end{align}
where $\eta$ is the Benzeggagh-Kenane interaction parameter.
The pure-mode jump components corresponding to fracture initiation and complete fracture are given by

\begin{align}
  &\jump{u}^0_n = \frac{f_{n}}{K_n}, \qquad \,~\jump{u}^f_n = \frac{2G_{Ic,d}}{f_{n}} \\
  &\jump{u}^0_{sh} = \frac{f_{sh}}{K_{sh}}, \qquad \jump{u}^f_{sh} = \frac{2G_{IIc,d}}{f_{sh}} 
  \label{eq:initiation-final-jumps}
\end{align}

An energy-based damage variable $\Dam$ is introduced as the state variable, which is defined as the ratio of dissipated energy $G_d$ over the critical mixed-mode energy release rate $G_c$

\begin{equation}
  \Dam \equiv \frac{G_d}{G_c} = \frac{\Delta-\Delta_f}{\Delta_f - \Delta_0}
  \label{eq:bigd-def}
\end{equation}
and can only increase in \emph{pseudo} time $t$, such that for time step $n$ ($t=t_n$):

\begin{equation}
  \Dam(t_n) = \max \limits_{0 \leq \tau \leq t_n} \big(\Dam(\tau)\big)
  \label{eq:smalld-int}
\end{equation}

The stiffness-based damage variable $d$ in \cref{eq:traction-update} is related to the energy-based damage variable through

\begin{equation}
  d= 1-\frac{(1-\Dam)\Delta_0}{\Dam \Delta_f + (1-\Dam) \Delta_0}
  \label{eq:bigd-smalld-rel}
\end{equation}

\subsection*{\Davila's fatigue damage formulation}
 
\begin{figure}
  \centering
  \begin{tikzpicture}[>=stealth]
    \node(aa) at (9.5,1.0) [rectangle] {\def\svgwidth{0.5\textwidth}\scalebox{0.85}{
\begingroup%
  \makeatletter%
  \providecommand\color[2][]{%
    \errmessage{(Inkscape) Color is used for the text in Inkscape, but the package 'color.sty' is not loaded}%
    \renewcommand\color[2][]{}%
  }%
  \providecommand\transparent[1]{%
    \errmessage{(Inkscape) Transparency is used (non-zero) for the text in Inkscape, but the package 'transparent.sty' is not loaded}%
    \renewcommand\transparent[1]{}%
  }%
  \providecommand\rotatebox[2]{#2}%
  \newcommand*\fsize{\dimexpr\f@size pt\relax}%
  \newcommand*\lineheight[1]{\fontsize{\fsize}{#1\fsize}\selectfont}%
  \ifx\svgwidth\undefined%
    \setlength{\unitlength}{233.68890958bp}%
    \ifx\svgscale\undefined%
      \relax%
    \else%
      \setlength{\unitlength}{\unitlength * \real{\svgscale}}%
    \fi%
  \else%
    \setlength{\unitlength}{\svgwidth}%
  \fi%
  \global\let\svgwidth\undefined%
  \global\let\svgscale\undefined%
  \makeatother%
  \begin{picture}(1,0.06879384)%
    \lineheight{1}%
    \setlength\tabcolsep{0pt}%
    \put(0,0){\includegraphics[width=\unitlength,page=1]{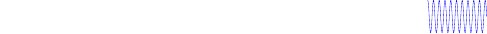}}%
    \put(0.83910602,0.01940553){\color[rgb]{0,0,0}\makebox(0,0)[lt]{\lineheight{1.25}\smash{\begin{tabular}[t]{l}$\sigma$\end{tabular}}}}%
    \put(0,0){\includegraphics[width=\unitlength,page=2]{1Dbar.pdf}}%
    \put(0.1383028,0.01866251){\color[rgb]{0,0,0}\makebox(0,0)[lt]{\lineheight{1.25}\smash{\begin{tabular}[t]{l}$\sigma$\end{tabular}}}}%
    \put(0,0){\includegraphics[width=\unitlength,page=3]{1Dbar.pdf}}%
  \end{picture}%
\endgroup%
}};
      \node at (10,-1.5) [rectangle] {\def\svgwidth{1.0\columnwidth}\scalebox{0.85}{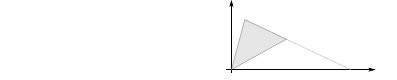}};
      \node (a) at (9.49,0.9) {};
      \node (b) at (11.5,-0.5) {};
      \node (c) at (11.9,-1.4) {};
      \node (d) at (12.9,-1.0) {\small$G_{d}$};
      \draw [->,blue] (a.north) to [out=230, in=90] (b.north);
      \draw [-,black] (c.north) to [out=30, in=180] (d.west);
  \end{tikzpicture}
  \caption{\Davila's fatigue cohesive zone model. The evolution of damage variable $\Dam$ is such that at constant stress and mode-mixity the time to failure $N_{\mathrm{fail}}$ matches with an S-N curve}
  \label{fig:fatigue-czm}
\end{figure}

\begin{figure}
  \hspace{10em}
  \begin{tikzpicture}
    \node at (0,0) {\def\svgwidth{0.65\columnwidth}\scalebox{1.00}{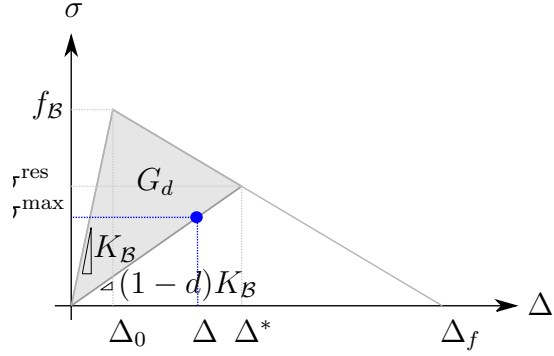}};
  \end{tikzpicture}
  \caption{Nomenclature of the fatigue CZM. The fatigue traction-separation response ($\textcolor{blue}{\bullet}$) is inside the quasi-static envelope}
  \label{fig:nomenclature-fczm}
\end{figure}

In \Davila's fatigue CZM, the number of cycles to failure of a 1D bar with a single crack and cyclic load matches with an S-N curve (see \cref{fig:fatigue-czm}). This is achieved by reducing the stiffness at constant applied stress until the traction-separation response reaches the quasi-static softening line, which marks failure of the material point. The evolution of the energy-based damage variable during fatigue $\Dam_f$ is described with the following (nonlinear) differential equation
\begin{equation}
  \dv{\Dam_f}{N} = f_{\Dam}(\Delta, \Delta^*, \Dam) 
  \label{eq:bigd-evol}
\end{equation}
where $\Delta^*$ is the reference displacement, which is the displacement corresponding to the residual traction (see \cref{fig:nomenclature-fczm}) and can be computed as

\begin{equation}
  \Delta^* = \Dam(\Delta_f - \Delta_0) + \Delta_0
  \label{eq:refdisp}
\end{equation}

The quasi-static damage $\Dam_s$ is computed as

\begin{equation}
  \Dam_s = \frac{\Delta - \Delta_0 }{ \Delta_f - \Delta_0 }
  \label{eq:dam-static}
\end{equation}
and the updated damage is determined as the maximum of the static and the fatigue damage

\begin{equation}
  \Dam = \max\big(\Dam_s, \Dam_f\big)
  \label{eq:dam-static-or-fatigue}
\end{equation}
to ensure that the traction-opening response during fatigue loading is inside the quasi-static envelope.

In \cite{Davila2020Nasa}, several fatigue damage functions were proposed and compared. It was shown that the so-called CF20 damage function

\begin{align}
 &f^{\mathrm{CF20}}_{\Dam} = \frac{1}{\gamma} \frac{(1-\Dam)^{\beta-p}}{E^\beta (p+1)}\bigg(\frac{\Delta}{\Delta^*}\bigg)^\beta
 \label{eq:CF20} 
\end{align}
gave the most satisfactory results. Here, $\gamma$ is the number of cycles to failure at the endurance limit (which is usually set to $10^{7}$ cycles), $p$ can be calibrated such that the propagation rates in the simulation match with (available) Paris curves \cite{Davila2020Nasa} and $\beta$ is the exponent in the S-N curve, computed as
\begin{equation}
  \beta = \frac{-7 \eta}{log E}
\end{equation}
where $\eta$ is a \emph{brittleness} parameter that can take into account the low-cycle fatigue response in the S-N curve. 

For a given stress ratio $R\equiv \sigma^{\mathrm{min}}/\sigma^{\mathrm{max}}$, the relative endurance limit $E$, defined as the ratio of equivalent endurance limit $\sigma_{\mathrm{end}}$ and mode-dependent static strength $f_{\B}$, is computed from the relative endurance limit $\epsilon$ (at full load reversal $R=-1$) with the Goodmann diagram:
\begin{equation}
  E = \frac{2 C_l \epsilon}{C_l\epsilon + 1 + R(C_l \epsilon - 1)}
  \label{eq:endurance-R-B-dependence}
\end{equation}
where $C_l$ is an empirical relation which takes into account the effect of mode-mixity \cite{juvinallFundamentalsMachineComponent2012}:

\begin{equation}
  C_l = 1-0.42 \B
  \label{eq:juvinall-relation}
\end{equation}

In summary, the input fatigue model parameters for CF20 are $\eta, \epsilon$ and $p$.
 
\subsection*{Implicit fatigue damage update}
In order to compute the damage at current \emph{pseudo} time $t_n$, the damage rate function in \cref{eq:bigd-evol} must be integrated, which is mathematically expressed as

\begin{equation}
  \Dam_{f}^{t_n} = \Dam^{t_{n-1}} + \int^{t_n}_{t_{n-1}} f_{\Dam}(\Delta, \Delta^*, \Dam)\, \mathrm{d}N
\end{equation}

\citeauthor{Davila2020}~\cite{Davila2020} used an Euler forward (explicit) time integration scheme where the integral is approximated as $\Delta N f_\Dam^{(n-1)}$, with $f_\Dam^{(n-1)}$ representing $f_\Dam$ evaluated at $t_{n-1}$. In this work, the damage at current time step $n$ (corresponding to time  $t_n$) is computed with the generalized trapezoidal rule:
\begin{equation}
  \Dam^{(n)}_{f} = \Dam^{(n-1)} + \Delta N \Big[ (1-\theta) f_{\Dam}^{(n-1)} + \theta f_{\Dam}^{(n)} \Big]
  \label{eq:implicit-update}
\end{equation}
with parameter $\theta \in \big(0,1\big]$.\footnote{Note that for $\theta = 0$, the formulation reduces to an Euler forward scheme} Through \cref{eq:CF20}, $f_\Dam^{(n)}$ requires $\Dam_f^{(n)}$, making the update with \cref{eq:implicit-update} implicit. The resulting nonlinear equation can be solved at \emph{local} integration point level, \eg with Newton's method by performing iterations until a \emph{local} convergence criterion is met.

\subparagraph{Remark}
  The right-hand side (RHS) of \cref{eq:implicit-update} depends on the step size $\Delta N$. When the step size is too large and a material point completely fails within the time step, no solution exists for \cref{eq:implicit-update} (see \cref{fig:residual}). This issue is circumvented by setting $\Dam_f=1$ when this occurs.

\begin{figure}
  \centering
  \begin{tikzpicture}
    \node[] at (0,0)
    {
      \input{figures/dcb-residual.tex}
    };
  \end{tikzpicture}
  \caption{Right-hand side of \cref{eq:implicit-update} as a function of damage $\Dam_{f}^{(n)}$ (curves in \emph{gray}) at constant equivalent displacement $\Delta$ while increasing the cycle increment $\Delta N$ starting from $\Delta N=0$ (horizontal \emph{gray} line). For a given $\Delta N$, the solution of \cref{eq:implicit-update} is the intersection with the \emph{blue} line $\mathrm{RHS}=\Dam_{f}^{(n)}$. The damage value $\Dam_0$ of the previous \emph{pseudo} time step $n$ is indicated with $\textcolor{red}{\bullet}$. When the cycle increment is too large, there exists no intersection for $\Dam_{f}^{(n)}\in\big[\Dam_0,1\big)$ (curves in \emph{red}).}
  \label{fig:residual}
\end{figure}

\subsection*{Consistent linearization of the traction update}

In \cite{Davila2020} and \cite{Turon2018b}, a numerical tangent stiffness based on finite differences was used to approximate the tangent of the static cohesive relation and the extension with fatigue damage. However, the accuracy of approximating the tangent and robustness depends on the choice of the perturbation and is case-dependent. In addition, the traction update needs to be performed for each perturbation of the displacement jump component, which may not be computationally efficient. 

In order to improve the efficiency and robustness, which is crucial in full-laminate analyses, a consistent tangent stiffness matrix is derived below for the static cohesive zone model \cite{Turon2018b} and the fatigue damage extension \cite{Davila2020}, including the improved (implicit) fatigue damage update presented in this work.

The traction at current time $n$ is written as a function of displacement jump $\dispjump$, current damage $\Dam$ and cycle jump $\Delta N$
\begin{equation}
  \textbf{t} = \hat{\textbf{t}}(\dispjump,\Dam(\dispjump, \Delta N ))
\end{equation}

The linearized traction update can be expressed as

\begin{equation}
  \delta \textbf{t} = \underbrace{\Bigg(\pdv{\hat{\textbf{t}}(\dispjump, \Dam(\dispjump, \Delta N ))}{\dispjump} \Bigg)_{(n)}}_{\textbf{D}} \delta \dispjump
\end{equation}
where it is explicitly indicated that the time $n$ is kept constant. The material tangent $\textbf{D}$ can be identified in this expression as the linear operator that maps an iterative-increment of the displacement jump to an iterative-increment of the traction vector. 

By performing partial differentiation and after some re-arrangement, the expression of the material tangent becomes
\begin{align}
  \textbf{D} =\underbrace{(\textbf{I} - d\textbf{P})\textbf{K}}_{\mathrm{secant~stiffness}} -  \textbf{P}\textbf{K}\jump{\textbf{u}} \pdv{d^{^\mathrm{T}}}{\dispjump}
  \label{eq:mat-tang-expanded}
\end{align}

Differentiation of the last term in \cref{eq:mat-tang-expanded} and applying the chain rule gives

\begin{align}
  \pdv{d}{\dispjump} &= \pdv{d}{\Dam}\pdv{\Dam}{\dispjump} + \pdv{d}{\Delta_0} \pdv{\Delta_0}{\dispjump}
  + \pdv{d}{\Delta_f}\pdv{\Delta_f}{\dispjump} 
  \label{eq:dd_djump}
\end{align}
where the partial derivatives $\hpdv{\Delta_0}{\dispjump}$ and $\hpdv{\Delta_f}{\dispjump}$ can be further expanded:

\begin{align}
  \pdv{\Delta_0}{\dispjump} &= \pdv{\Delta_0}{\B} \pdv{\B}{\dispjump} + \pdv{\Delta_0}{K_{\B}}\pdv{K_{\B}}{\B}\pdv{\B}{\dispjump}
  \label{eq:delta0_du} \\
  \pdv{\Delta_f}{\dispjump} &= \pdv{\Delta_f}{\B} \pdv{\B}{\dispjump}  + \pdv{\Delta_f}{K_{\B}}\pdv{K_{\B}}{\B}\pdv{\B}{\dispjump} + \pdv{\Delta_f}{\Delta_0} \pdv{\Delta_0}{\dispjump}
  \label{eq:deltaf_du}
\end{align}

The partial derivatives in \cref{eq:dd_djump,eq:delta0_du,eq:deltaf_du} can be computed with quantities already obtained from the traction update algorithm presented before:

\begin{align}
  \pdv{d}{\Dam} &= \frac{\Delta_0 \Delta_f}{\big[(\Dam-1)\Delta_0 - \Dam \Delta_f \big]^2} \\
  \pdv{d}{\Delta_0} &= \frac{(\Dam-1)\Dam \Delta_f}{\big[(\Dam-1)\Delta_0 - \Dam \Delta_f \big]^2} \\
  \pdv{d}{\Delta_f} &= \frac{(1-\Dam)\Dam \Delta_0}{\big[(\Dam-1)\Delta_0 - \Dam \Delta_f \big]^2} \\
  \pdv{\Delta_0}{\B} &= \frac{K_{sh}(\jump{u}^0_{sh})^2 - K_n(\jump{u}^0_n)^2}{2 \Delta_0 K_{\B}} \B^{\eta-1}\eta \\
  \pdv{\Delta_0}{K_{\B}} &= - \frac{\Delta_0}{2K_{\B}} \\
  \pdv{\Delta_f}{\B} &= \frac{K_{sh}\jump{u}^0_{sh} \jump{u}^f_{sh} -K_n \jump{u}^0_{n} \jump{u}^f_{n}}  {\Delta_0 K_{\B}} \B^{\eta-1}\eta\\
  \pdv{\Delta_f}{K_{\B}} &= -\frac{\Delta_f}{K_{\B}}\\
  \pdv{K_{\B}}{\B} &= -K_n+K_{sh}\\
  \pdv{\B}{\dispjump} &= \frac{2 K_n K_{sh}}{\Big(K_n \mc{\jump{u}_n}^2 + K_{sh} \jump{u}_{sh}^2 \Big)^2} \notag\\ 
  &\qquad\Bigg[-(\jump{u}_{sh})^2 \mc{\jump{u}_n}, \jump{u}_{s1} \mc{\jump{u}_n}^2, \jump{u}_{s2} \mc{\jump{u}_n}^2 \Bigg]^\mathrm{T}\\
  \pdv{\Delta_f}{\Delta_0} &= - \frac{\Delta_f}{\Delta_0}
\end{align}  

During loading, $\hpdv{\Dam}{\dispjump}$ is non-zero at current time $n$ and global iteration $j$. Due to the $\mathrm{max}$ operator in \cref{eq:dam-static-or-fatigue}, the term $\hpdv{\Dam}{\dispjump}$ is not continuous which requires considering each case (static and fatigue loading) separately.

\subparagraph{Quasi-static loading}
When quasi-static damage is larger than fatigue damage, the term $\hpdv{\Dam}{\dispjump}$ is obtained by partial differentiation of the static damage (\cref{eq:dam-static}):
\begin{equation}
  \pdv{\Dam}{\dispjump} = \pdv{\Dam_s}{\Delta} \pdv{\Delta}{\dispjump} + \pdv{\Dam_s}{\Delta_0}\pdv{\Delta_0}{\dispjump} + \pdv{\Dam_s}{\Delta_f}\pdv{\Delta_f}{\dispjump}
\end{equation}
with

\begin{align}
  \pdv{\Delta}{\dispjump} &=\Big(K_n^2 \mc{\jump{u}_n}^2 + K_{sh}^2 \jump{u}_{sh}^2\Big)^{-3/2}\notag\\
  &\qquad\begin{bmatrix}
  K_n \mc{\jump{u}_n} \big[ K_n^2 \mc{\jump{u}_n}^2 + (2 K_{sh}^2 - K_n K_{sh}\big) \jump{u}_{sh}^2 \big] \\  
K_{sh}\jump{u}_{s1} \big( 2 K_n^2 \mc{\jump{u}_n}^2 - K_n K_{sh} \mc{\jump{u}_n}^2 + K_{sh}^2 \jump{u}_{sh}^2\big) \\ 
K_{sh}\jump{u}_{s2} \big( 2 K_n^2 \mc{\jump{u}_n}^2 - K_n K_{sh} \mc{\jump{u}_n}^2 + K_{sh}^2 \jump{u}_{sh}^2\big)  
\end{bmatrix}
\\
  \pdv{\Dam_s}{\Delta} &= \frac{1}{\Delta_f - \Delta_0} \\
  \pdv{\Dam_s}{\Delta_0} &= \frac{\Delta - \Delta_f}{(\Delta_f-\Delta_0)^2} \\
  \pdv{\Dam_s}{\Delta_f} &= \frac{\Delta_0 - \Delta}{(\Delta_f-\Delta_0)^2} 
\end{align}

\subparagraph{Fatigue loading}
When fatigue damage is larger than quasi-static damage, the term must be determined differently. The implicit damage update in \cref{eq:implicit-update} can be recast in residual form as

\begin{equation}
  r =  \Dam^{(n)} - \Dam^{(n-1)} - \Delta N \Big[ (1-\theta)f_{\Dam}^{(n-1)} + \theta f_{\Dam}^{(n)} \Big]
  \label{eq:residual}
\end{equation}

The residual is a function of independent variables $\Dam$ and $\dispjump$. Therefore, the variation of the residual is expressed as
\begin{equation}
\delta r = \Bigg(\pdv{r}{\Dam}\Bigg)_{\dispjump} \delta \Dam + \Bigg(\pdv{r}{\dispjump}\Bigg)_{\Dam} \delta \dispjump
\label{eq:delta-res}
\end{equation}

The \emph{local} damage state at current time $n$ is obtained in every \emph{global} iteration by iteratively solving for the residual to be zero (within a sufficiently small tolerance). Therefore, the variation of the \emph{locally converged} residual does not change between \emph{global} iterations:
\begin{equation}
  \delta r = 0
\end{equation}
which is a consistency condition. Similar to what is done for plasticity models with return mapping algorithms, this consistency condition can be used to obtain a relation between the variation of the displacement jump and the \emph{locally converged} damage variable $\Dam$ from which the derivative of the damage with respect to displacement jump can be identified:

\begin{equation}
\delta \Dam= \underbrace{- \frac{\big(\pdv{r}{\dispjump}\big)_{\Dam}} {\big(\pdv{r}{\Dam}\big)_{\dispjump}}}_{\pdv{\Dam}{\dispjump}} \delta \dispjump
\end{equation}

Applying the chain-rule gives the expressions for the partial derivatives

\begin{align}
&  \bigg(\pdv{r}{\Dam}\bigg)_{\dispjump}  =  \pdv{r}{\Dam} + \pdv{r}{\Delta^*}\pdv{\Delta^*}{\Dam}  
\label{eq:dr_dD_constdu}\\
&  \bigg(\pdv{r}{\dispjump}\bigg)_{\Dam} = \pdv{r}{\Delta} \pdv{\Delta}{\dispjump} + \pdv{r}{\Delta^*}\ \pdv{\Delta^*}{\dispjump} + \sum^{N\fatprops}_{i=1}\pdv{r}{\fatprops_i}\pdv{\fatprops_i}{\dispjump}
\label{eq:dr_du_constD}
\end{align}
where the last terms after the summation are the partial derivatives of the parameter functions that depend on the displacement jump (for CF20: $\mathcal{P}=\{E,\beta, p\}$). Applying the chain rule to the fourth term in \cref{eq:dr_du_constD} gives

\begin{align}
  &\pdv{\Delta^*}{\dispjump} = \pdv{\Delta^*}{\Delta_0}\pdv{\Delta_0}{\dispjump} + \pdv{\Delta^*}{\Delta_f} \pdv{\Delta^f}{\dispjump}
\label{eq:partial-res-Delta}
\end{align}
where $\hpdv{\Delta_0}{\dispjump}$ and $\hpdv{\Delta_f}{\dispjump}$ are derived earlier (see \cref{eq:delta0_du,eq:deltaf_du}). Performing the differentiation of the other derivatives in \cref{eq:dr_dD_constdu,eq:dr_du_constD,eq:partial-res-Delta} gives
\begin{align}
 &\pdv{r}{\Delta} = -\Delta N \theta \pdv{f_{\Dam}}{\Delta}\\
 &\pdv{r}{\Delta^*} = -\Delta N \theta \pdv{f_{\Dam}}{\Delta^*} \\
 &\pdv{r}{\Dam} = 1 - \Delta N \theta \pdv{f_{\Dam}}{\Dam}\\
 &\sum^{N\fatprops}_{i=1}\pdv{r}{\fatprops_i}\pdv{\fatprops_i}{\dispjump} = -\Delta N \theta \Bigg[\sum^{N\fatprops}_{i=1} \pdv{f_{\Dam}}{\fatprops_i} \pdv{\fatprops_i}{\B}\Bigg] \pdv{\B}{\dispjump} \label{eq:sumP}\\
 &\pdv{\Delta^*}{\Dam} = \Delta_f - \Delta_0 \\ 
 &\pdv{\Delta^*}{\Delta_f} = \Dam \\
 &\pdv{\Delta^*}{\Delta_0} = 1-\Dam 
\end{align}
where, for CF20 with $\mathcal{P}=\{E,\beta, p\}$, the term in \cref{eq:sumP} can be further expanded:
\begin{align}
 &\sum^{N\fatprops}_{i=1} \pdv{f_{\Dam}}{\fatprops_i} \pdv{\fatprops_i}{\B} = \Bigg[\pdv{f_{\Dam}}{E} + \Bigg(\pdv{f_{\Dam}}{\beta} + \textcolor{black}{\pdv{f_{\Dam}}{p}\pdv{p}{\beta}} \Bigg) \pdv{\beta}{E} \Bigg] \pdv{E}{C_l}\pdv{C_l}{\B} \\
\end{align}

with
\begin{align}
  &\pdv{f_{\Dam}}{\Delta} =  \frac{\beta}{\Delta} f_{\Dam}\\
  &\pdv{f_{\Dam}}{\Delta^*} = -\frac{\beta}{\Delta^*} f_{\Dam}\\
  &\pdv{f_{\Dam}}{\Dam} =  \frac{p-\beta}{1-\Dam} f_{\Dam} \\
  &\pdv{f_{\Dam}}{E} =  - \frac{\beta}{E} f_{\Dam}\\
  &\pdv{f_{\Dam}}{\beta} =  \big[ \mathrm{ln}\Big(\frac{\Delta}{\Delta^*}\Big) + \mathrm{ln}(1-\Dam) - \mathrm{ln}(E) \big] f_{\Dam} \\
  &\textcolor{black}{\pdv{f_{\Dam}}{p} = -\frac{f_{\Dam}}{p+1} \big[1+(p+1) \mathrm{ln}(1-\Dam)  \big]}\\
  &\pdv{E}{C_l} = \frac{2 \epsilon(1-R)}{\big[C_l\epsilon(R + 1) - R + 1\big]^2} \\
  &\pdv{C_l}{\B} = -0.42 \\
  &\pdv{\beta}{E} = \frac{7\eta \ln{(10)}}{\ln{(E)}^2 E} \\
  &\textcolor{black}{\pdv{p}{\beta} = 1}
\end{align}

\subsection*{Phantom node version of XFEM}
\begin{figure}
  \centering
  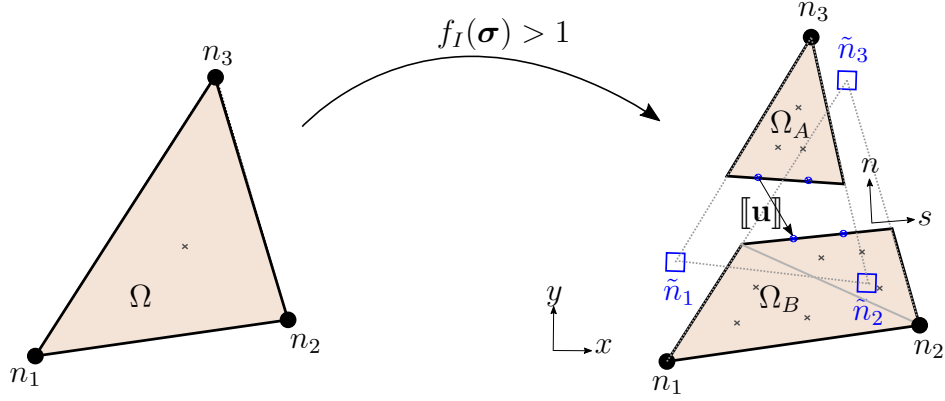
  \caption{Crack insertion in XFEM element. When the stress in a bulk integration point (\emph{left}) satisfies the insertion criterion, a cohesive segment is inserted (\emph{right})}
  \label{fig:cracked-elem}
\end{figure}

The fatigue CZM with the improved damage update is combined with the phantom node version of XFEM \cite{Hansbo2004}. When a certain stress criterion in a bulk integration point is reached, a discontinuity is inserted in the displacement field of the element (see \cref{fig:cracked-elem}). In order to include microstructural information of cracks propagating in the fiber direction, the normal $\textbf{n}$ of the crack is fixed in the direction perpendicular to the fibers following Van der Meer and Sluys \cite{Meer2009}. 

A discontinuity in the displacement field is achieved by duplicating the original element and expressing the displacement field in terms of the independent displacement fields of the two overlapping sub-elements:

\begin{equation}
\textbf{u}(\textbf{x}) =  
\left\{
\begin{array}{ll}
  \textbf{N}(\textbf{x})\textbf{u}_{A},~\textbf{x}\in \Omega_{A} \\
  \textbf{N}(\textbf{x})\textbf{u}_{B},~\textbf{x}\in \Omega_{B} \\
\end{array} 
\right.
\end{equation}
where $\textbf{u}_{A}$ and $\textbf{u}_{B}$ are the vectors containing the nodal DOFs of the sub-elements with original nodes and \emph{phantom nodes}. The connectivity of the sub-elements is given as
\begin{align}
  &\Omega^n_A = \{n_1,n_2,\textcolor{blue}{\tilde{n}_3}\} \\
  &\Omega^n_B = \{\textcolor{blue}{\tilde{n}_1},\textcolor{blue}{\tilde{n}_2},n_3\}
\end{align}

The displacement jump vector along the cohesive segment $\Gamma_d$ is defined as

\begin{equation}
\dispjump (\textbf{x}) = \textbf{N}(\textbf{x})(\textbf{u}_{A}-\textbf{u}_{B}),~\textbf{x} \in \Gamma_d 
\end{equation}

\subsection*{Fatigue XFEM insertion criterion}
\label{sec:fatigue-xfem-insertion-criterion}
In order to allow for matrix cracks to initiate at arbitrary locations, an XFEM fatigue crack insertion criterion is presented in the following. In the fatigue CZM, fatigue damage already accumulates before the static initiation stress is reached, provided that the stress is above the endurance limit. Therefore, the endurance limit is a natural choice to define as the moment for insertion of cohesive crack segments, which is consistent with the fatigue damage formulation. 

\begin{figure}
  \centering
  \begin{tikzpicture}
  \node[] at (0,0)
    {
      \input{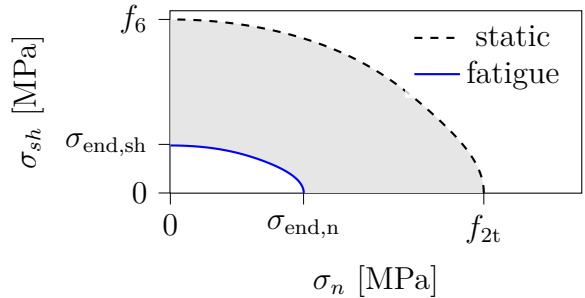}
    };
  \end{tikzpicture}
  \caption{Fatigue and static damage initiation surfaces in stress-space}
  \label{fig:contours}
\end{figure}

The \emph{relative} endurance limit depends on the \emph{local} stress ratio $R$ and mode-mixity $\B$ via \cref{eq:endurance-R-B-dependence,eq:juvinall-relation}. The endurance limit can be computed with the relative endurance limit and the static equivalent strength as 

\begin{equation}
\sigma_{\mathrm{end}} = E f_\B
\label{eq:endurance-limit}
\end{equation}
where the static mode-dependent strength $f_\B$ is related to the mode-dependent dummy stiffness (\cref{eq:mode-dependent-dummy}) and the equivalent initiation jump (\cref{eq:initiation-jump}):

\begin{equation}
  f_\B = K_\B \Delta_0 
  \label{eq:mode-dependent-strength-zero-damage}
\end{equation}
Substituting the pure-mode initiation displacement components (\cref{eq:initiation-final-jumps}) in the expression for the equivalent fracture initiation jump (\cref{eq:initiation-jump}) and substituting the result, together with the expression for the mode-dependent dummy stiffness (\cref{eq:mode-dependent-dummy}), in \cref{eq:mode-dependent-strength-zero-damage} gives

\begin{equation}
  f_\B = \sqrt{(K_n(1-\B)+\B K_{sh}) \big[f^2_{n}/K_n + ( f_{sh}^2/K_{sh} - f^2_{n}/K_n) \B^{\eta}\big]}
  \label{eq:mode-dependent-strength-expanded}
\end{equation}
which is an expression in terms of input material model parameters and mode-mixity only. The material model parameters are readily available in a bulk integration point. However, in the absence of a cohesive segment in the XFEM element before crack insertion, \cref{eq:mode-mixity-def} cannot be used to compute the displacement-based mode-mixity $\B$. Since the normal $\textbf{n}$ is fixed in the direction of the fibers and known before crack insertion, the traction in each bulk element can be computed with the bulk stress using $\textbf{t}=\bm{\sigma} \textbf{n}$. By using the fact that before crack insertion damage is zero ($d=0 \rightarrow t_n = K_n \llbracket u \rrbracket_n, t_{sh} = K_{sh} \llbracket u\rrbracket_{sh}$), the  mode-mixity can be computed in each integration point of a bulk element:

\begin{equation}
  \B = \frac{t_{sh}/K_{sh}}{\mc{t_n} /K_n + t_{sh} /K_{sh}}
  \label{eq:mode-mixity-def-stress-based}
\end{equation}
from which the endurance limit and the mode-dependent strength can be computed with \cref{eq:endurance-limit,eq:mode-dependent-strength-zero-damage,eq:mode-dependent-strength-expanded}. The equivalent stress in the bulk integration point is computed with \cref{eq:sigma-equiv}.
Finally, the XFEM fatigue crack insertion criterion is defined as
\begin{equation}
  f_I(\bm{\sigma})\equiv \frac{\sigma(\bm{\sigma})}{\sigma_{\mathrm{end}}(\bm{\sigma})} > 1.0
  \label{eq:failure-index-function}
\end{equation}
which represents a surface in stress-space when $f_I(\bm{\sigma})=1.0$ (see \cref{fig:contours}).

It should be emphesized that at the moment of crack insertion when \cref{eq:failure-index-function} is satisfied, fatigue damage in the newly inserted cohesive crack segment is zero. Only \emph{after} crack insertion, fatigue damage can accumulate according to the formulation presented in the previous section.

\subsection*{Shifted cohesive relation}
Cohesive XFEM segments are inserted on the fly when the stress criterion is reached. The static initiation stress used in this work is the B-K interpolation of the elastic stored energy \cite{Gonzalez2009} before the peak at zero damage (see \cref{eq:mode-dependent-strength-zero-damage}).
For fatigue damage, the fatigue crack insertion criterion presented in the prevous section is used. 

At the time of insertion, the stress in the material is non-zero. Therefore, following \cite{VanderMeer2012a}, Hille's approach is used \cite{Hille2009} where a shift is applied such that the traction for zero opening in the cohesive segment is in equilibrium with the stress in the bulk $\bm{\sigma}$ before and after insertion. The shift at the moment of crack insertion is computed as

\begin{equation}
  \dispjump_{\mathrm{shift}} = \textbf{K}^{-1} \bm{\sigma} \textbf{n}
\end{equation}

\begin{figure}
  \centering
  {\def\svgwidth{0.8\columnwidth}{\scalebox{1.0}{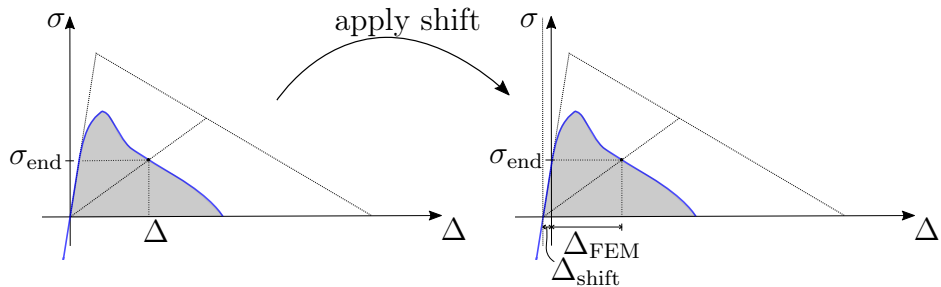}}}
  \caption{Shifted fatigue cohesive zone model}
  \label{fig:shifted-fatigue-czm}
\end{figure}

The updated traction at current time $n$ is computed with \cref{eq:traction-update} where the displacement jump is the sum of the jump passed to the integration point $\dispjump_{\mathrm{FEM}}$ and the shift $\dispjump_{\mathrm{shift}}$ (see \cref{fig:shifted-fatigue-czm}). 

\subsection{Adaptive cycle jumping}
\label{sec:adaptive-stepping}

The cycle increment during fatigue loading is determined with a measure based on the number of \emph{global} Newton-Raphson iterations $N_{\mathrm{iter}}$, following a strategy similar to \cite{verhooselDissipationbasedArclengthMethod2009} as previously applied in static loading \cite{Meer2009}. The cycle increment $\Delta N$ for the next time step $n+1$ is computed from the current (converged) time step $n$ as

\begin{equation}
  \Delta N^{(n+1)} = C^{-(\frac{n_{\mathrm{iter}}-n_{\mathrm{iter}}^{\mathrm{opt}}}{\xi})} \Delta N^{(n)}
  \label{eq:adaptive-step}
\end{equation}
where $C$, $\xi$ and $N_{\mathrm{iter}}^{\mathrm{opt}}$ are model parameters. If convergence is not reached within a specified maximum number of iterations $n_{\mathrm{iter}}^{\mathrm{max}}$, the step is cancelled and restarted with a reduced cycle increment $\Delta N^{(n)} \leftarrow c_{\mathrm{red}} \Delta N^{(n)}$.

\section{Examples}
First a simple 1D case is simulated with a single XFEM element and shifted cohesive relation in order to verify the presented modeling approaches and to demonstrate the improved accuracy with the implicit fatigue damage update. As a second example, a DCB test is simulated where the importance of an implicit scheme for integrating the damage rate function is highlighted. In the last example, an open-hole [$\pm45$]$_s$-laminate simulation is shown to demonstrate the capabilities of the numerical framework in simulating the complex interaction of matrix cracking and interface delamination under fatigue loading.
\subsection{Example A: Single XFEM cohesive element test}
\begin{figure}
  \centering
   {\def\svgwidth{0.6\columnwidth}{\scalebox{1.0}{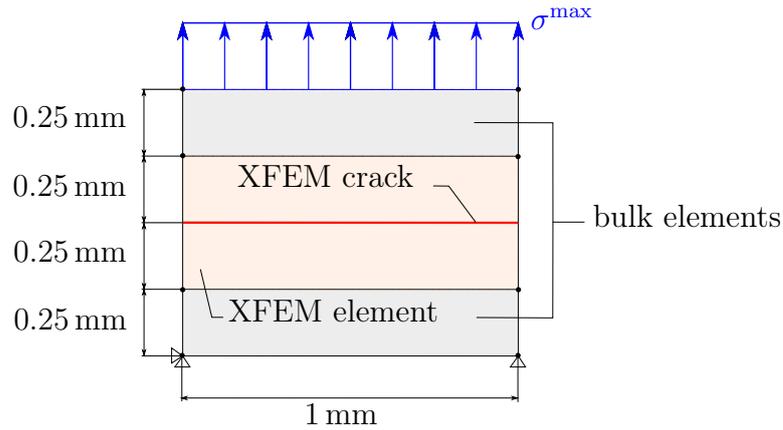}}}
  \caption{Specimen dimensions of single element test}
  \label{fig:element-dimensions}
\end{figure}

Firstly, a simple case under uniaxial tension is simulated (see \cref{fig:element-dimensions}). The maximum applied stress level is $\sigma^{\mathrm{\mathrm{max}}}=\SI{6}{MPa}$. The tensile strength, mode-I fracture energy and dummy stiffness are $f_t=\SI{10}{MPa}$, $G_{Ic}=0.1$ and $K = \SI{e4}{N/mm}$. The fatigue model parameters are $p=\beta$, $\eta=0.8$ and $\epsilon=0.2$. The XFEM crack is inserted in the middle element when the applied stress $\sigma^{\mathrm{max}}=0.2 f_t$.

For the case of constant tension, the damage evolution can be analytically derived as shown in \cite{Davila2020Nasa}. The time to failure is given with the following equation

\begin{equation}
    N_{\mathrm{fail}} = \bar{N}(\sigma^{\mathrm{max}}) =  \gamma E ^\beta \Big(\frac{\sigma^{\mathrm{max}}}{f_t}\Big)^{-\beta} \Big[1-\Big(\frac{\sigma^{\mathrm{max}}}{f_t}\Big)^{p+1}\Big]
    \label{eq:Nfail}
\end{equation}
where $\gamma=10^7$ (see \cref{eq:CF20}).

In the case of the maximum stress level considered in this example, the damage state at which the material point is considered to have failed is equal to $\Dam_{\mathrm{fail}}=1-\hfrac{\sigma^{\mathrm{max}}}{f_t}=0.4$.

\begin{figure}
\begin{subfigure}{0.5\textwidth}
  \begin{tikzpicture}
  \node[] at (0,0)
    {
      \input{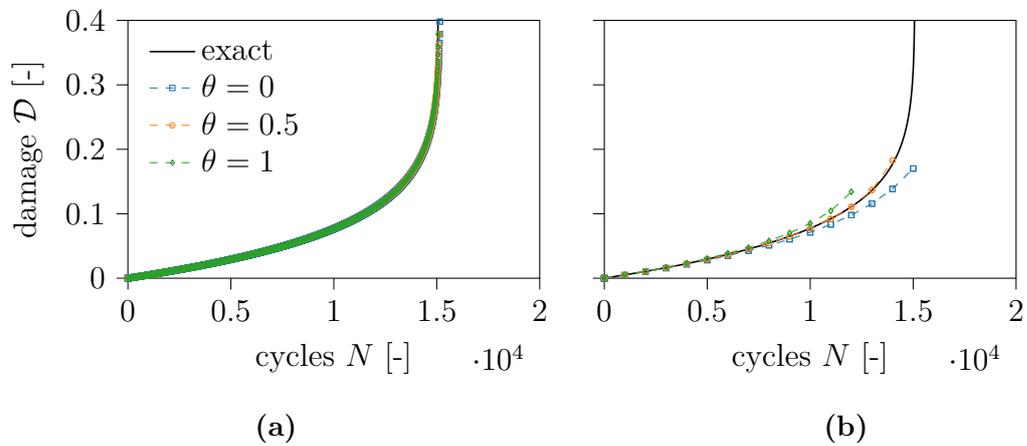}
    };
  \end{tikzpicture}
  \caption{}
  \label{fig:DvsN-dN-1000}
\end{subfigure}%
\begin{subfigure}{0.5\textwidth}
   \begin{tikzpicture}
  \node[] at (0,0)
    {
\begin{tikzpicture}

\definecolor{darkgray176}{RGB}{176,176,176}
\definecolor{darkorange25512714}{RGB}{255,127,14}
\definecolor{forestgreen4416044}{RGB}{44,160,44}
\definecolor{steelblue31119180}{RGB}{31,119,180}

\begin{axis}[
height=5cm,
minor xtick={},
minor ytick={},
scaled y ticks=manual:{}{\pgfmathparse{#1}},
tick align=outside,
tick pos=left,
width=7cm,
x grid style={darkgray176},
xlabel={cycles \(\displaystyle N\) [-]},
xmin=0, xmax=20000,
xtick style={color=black},
xtick={0,5000,10000,15000,20000},
y grid style={darkgray176},
ymin=0, ymax=0.4,
ytick style={color=black},
ytick={0,0.1,0.2,0.3,0.4},
yticklabels={}
]
\addplot [semithick, black]
table {%
0 0
1934.2666015625 0.0101009607315063
3636.51293945312 0.0202020406723022
5132.5986328125 0.0303030014038086
6445.72509765625 0.0404040813446045
7596.68603515625 0.0505050420761108
8604.0947265625 0.0606060028076172
9484.59375 0.0707070827484131
10253.044921875 0.0808080434799194
10922.705078125 0.0909091234207153
11505.3828125 0.101010084152222
12011.5859375 0.111111164093018
12450.6474609375 0.121212124824524
12830.8515625 0.13131308555603
13159.537109375 0.141414165496826
13443.2001953125 0.151515126228333
13687.5771484375 0.161616206169128
13897.732421875 0.171717166900635
14078.1259765625 0.181818246841431
14232.6806640625 0.191919207572937
14364.8427734375 0.202020168304443
14477.6328125 0.212121248245239
14573.6943359375 0.222222208976746
14655.337890625 0.232323169708252
14724.580078125 0.242424249649048
14783.17578125 0.252525210380554
14832.6494140625 0.26262629032135
14874.3251953125 0.272727251052856
14909.3486328125 0.282828330993652
14938.708984375 0.292929291725159
14963.2607421875 0.303030252456665
14983.73828125 0.313131332397461
15000.771484375 0.323232293128967
15014.9013671875 0.333333373069763
15026.58984375 0.34343433380127
15036.2294921875 0.353535413742065
15044.1572265625 0.363636374473572
15050.6552734375 0.373737335205078
15055.9658203125 0.383838415145874
15060.2900390625 0.39393937587738
15063.80078125 0.404040336608887
};
\addplot [steelblue31119180, dashed, mark=square, mark size=1, mark options={solid,fill opacity=0}]
table {%
0 0
0 0
0 0
0 0
0 0
0 0
0 0
0 0
0 0
0 0
0 0
0 0
0 0
0 0
1000 0.00488579273223877
1000 0.00488579273223877
2000 0.0100805759429932
2000 0.0100805759429932
3000 0.015627384185791
3000 0.015627384185791
4000 0.0215785503387451
4000 0.0215785503387451
5000 0.0279991626739502
5000 0.0279991626739502
6000 0.0349717140197754
6000 0.0349717140197754
7000 0.0426020622253418
7000 0.0426020622253418
8000 0.051030158996582
8000 0.051030158996582
9000 0.0604449510574341
9000 0.0604449510574341
10000 0.0711119174957275
10000 0.0711119174957275
11000 0.0834187269210815
11000 0.0834187269210815
12000 0.0979627370834351
12000 0.0979627370834351
13000 0.115731954574585
13000 0.115731954574585
14000 0.138530015945435
14000 0.138530015945435
15000 0.1701500415802
15000 0.1701500415802
};
\addplot [darkorange25512714, dashed, mark=o, mark size=1, mark options={solid,fill opacity=0}]
table {%
0 0
0 0
0 0
0 0
0 0
0 0
0 0
0 0
0 0
0 0
0 0
0 0
0 0
0 0
1000 0.00504553318023682
1000 0.00504553318023682
2000 0.0104340314865112
2000 0.0104340314865112
3000 0.016217827796936
3000 0.016217827796936
4000 0.0224627256393433
4000 0.0224627256393433
5000 0.0292526483535767
5000 0.0292526483535767
6000 0.0366973876953125
6000 0.0366973876953125
7000 0.0449442863464355
7000 0.0449442863464355
8000 0.0541983842849731
8000 0.0541983842849731
9000 0.0647568702697754
9000 0.0647568702697754
10000 0.0770771503448486
10000 0.0770771503448486
11000 0.0919196605682373
11000 0.0919196605682373
12000 0.110710024833679
12000 0.110710024833679
13000 0.136731028556824
13000 0.136731028556824
14000 0.182692050933838
14000 0.182692050933838
};
\addplot [forestgreen4416044, dashed, mark=diamond, mark size=1, mark options={solid,fill opacity=0}]
table {%
0 0
0 0
0 0
0 0
0 0
0 0
0 0
0 0
0 0
0 0
0 0
0 0
0 0
0 0
1000 0.00521647930145264
1000 0.00521647930145264
2000 0.0108151435852051
2000 0.0108151435852051
3000 0.0168602466583252
3000 0.0168602466583252
4000 0.0234355926513672
4000 0.0234355926513672
5000 0.0306508541107178
5000 0.0306508541107178
6000 0.0386556386947632
6000 0.0386556386947632
7000 0.0476617813110352
7000 0.0476617813110352
8000 0.0579851865768433
8000 0.0579851865768433
9000 0.0701303482055664
9000 0.0701303482055664
10000 0.0849902629852295
10000 0.0849902629852295
11000 0.104437947273254
11000 0.104437947273254
12000 0.134081959724426
12000 0.134081959724426
};
\end{axis}

\end{tikzpicture}
    };
  \end{tikzpicture}
  \caption{}
  \label{fig:DvsN-dN-10}
\end{subfigure}
\caption{Damage evolution with two different constant cycle increments $\Delta N$. \textbf{(a)} $\Delta N = 10$, \textbf{(b)} $\Delta N = 1000$}
\label{fig:DvsN-integration-comparison}  
\end{figure}

The damage evolution as a result of three different time integration parameter values $\theta \in \{0, 0.5, 1\}$ (see \cref{eq:implicit-update}), corresponding to Euler forward, trapezoidal rule and Euler backward, is shown in \cref{fig:DvsN-integration-comparison}. When the step size is sufficiently small, as in the case with $\Delta N =10$, all three methods show good correspondence with the exact analytical result. When the step size is increased to $\Delta N=1000$, the trapezoidal rule with $\theta = 0.5$ results in the most accurate response as it is second-order accurate, while Euler forward and Euler backward underestimate and overestimate the damage accumulation, respectively. Also note that with $\Delta N=1000$, the final damage $\Dam_{\mathrm{fail}}=0.4$ is not reached. This is caused by the fact that in the constant stress simulation, no equilibrium solution exists when $\Dam>\Dam_{\mathrm{fail}}$.

The simulations are also performed with the adaptive cycle jump scheme described in \cref{sec:adaptive-stepping}, where the cycle increment size $\Delta N$ is chosen based on the convergence characteristics from the previous \emph{pseudo} time step. The damage evolution is shown in \cref{fig:DvsN-adaptive}. The number of elapsed cycles in each \emph{pseudo} time step as a result of the accumulation of the (adaptive) cycle increments is shown in \cref{fig:ivsN}. It can be observed that the implicit damage update with adaptive stepping based on global convergence behavior is capable of tracing the full evolution of damage with high accuracy and efficient time stepping.

\begin{figure}
  \centering
  \begin{tikzpicture}
  \node[] at (0,0)
    {
\begin{tikzpicture}

\definecolor{darkgray176}{RGB}{176,176,176}

\begin{axis}[
height=5cm,
legend cell align={left},
legend style={
  fill opacity=0.8,
  draw opacity=1,
  text opacity=1,
  at={(0.03,0.03)},
  anchor=south west,
  draw=none
},
log basis x={10},
log basis y={10},
minor xtick={},
minor ytick={0.02,0.03,0.04,0.05,0.06,0.07,0.08,0.09,0.2,0.3,0.4,0.5,0.6,0.7,0.8,0.9,2,3,4,5,6,7,8,9,20,30,40,50,60,70,80,90,200,300,400,500,600,700,800,900},
tick align=outside,
tick pos=left,
width=7cm,
x grid style={darkgray176},
xlabel={cycles \(\displaystyle N\) [-]},
xmin=0.3141135959229, xmax=20000000,
xmode=log,
xtick style={color=black},
xtick={0.001,0.1,10,1000,100000,10000000,1000000000,100000000000},
y grid style={darkgray176},
ylabel={stress factor \(\displaystyle \Large(\sigma^{\mathrm{max}}/\sigma_\mathrm{c}\Large)\) [-]},
ymin=0.1, ymax=1.1,
ymode=log,
ytick style={color=black},
ytick={0.01,0.1,1,10,100}
]
\addplot [semithick, black, dashed]
table {%
88775086.72428 0.3
49576199.4555072 0.314285714285714
28412167.1539239 0.328571428571429
16673495.8112621 0.342857142857143
9999992.02897705 0.357142857142857
6119018.40631144 0.371428571428571
3814304.7032647 0.385714285714286
2418881.89640331 0.4
1558674.3469783 0.414285714285714
1019451.57694466 0.428571428571429
676119.711841004 0.442857142857143
454299.222142436 0.457142857142857
309010.128557812 0.471428571428571
212616.863447995 0.485714285714286
147885.720801053 0.5
103918.224545282 0.514285714285714
73730.7845737531 0.528571428571429
52792.3777054504 0.542857142857143
38128.4387379177 0.557142857142857
27764.3227106674 0.571428571428571
20375.253068875 0.585714285714286
15063.4846624279 0.6
11214.8226597554 0.614285714285714
8405.23061453441 0.628571428571429
6339.41125754061 0.642857142857143
4810.01514415183 0.657142857142857
3670.29397594033 0.671428571428571
2815.60235705473 0.685714285714286
2170.77734501751 0.7
1681.4572941626 0.714285714285714
1308.06334533535 0.728571428571429
1021.59536930171 0.742857142857143
800.674061579237 0.757142857142857
629.445325171801 0.771428571428571
496.085636612219 0.785714285714286
391.729186846085 0.8
309.693006450071 0.814285714285714
244.913972524615 0.828571428571429
193.537408582096 0.842857142857143
152.614791953249 0.857142857142857
119.880443374441 0.871428571428571
93.5857097483918 0.885714285714286
72.375222897058 0.9
55.1941115238219 0.914285714285714
41.2180984635053 0.928571428571429
29.800600564952 0.942857142857143
20.4325202249679 0.957142857142857
12.7115538876173 0.971428571428571
6.31866851719492 0.985714285714286
1 1
};
\addlegendentry{analytical}
\addplot [semithick, blue, mark=o, mark size=1.5, mark options={solid,fill opacity=0}, only marks]
table {%
147239 0.5
};
\addlegendentry{simulation}
\addplot [semithick, blue, mark=o, mark size=1.5, mark options={solid,fill opacity=0}, only marks, forget plot]
table {%
15011.9 0.6
};
\addplot [semithick, blue, mark=o, mark size=1.5, mark options={solid,fill opacity=0}, only marks, forget plot]
table {%
2165.94 0.7
};
\addplot [semithick, blue, mark=o, mark size=1.5, mark options={solid,fill opacity=0}, only marks, forget plot]
table {%
390.566 0.8
};
\addplot [semithick, blue, mark=o, mark size=1.5, mark options={solid,fill opacity=0}, only marks, forget plot]
table {%
71.6496 0.9
};
\addplot [semithick, blue, mark=o, mark size=1.5, mark options={solid,fill opacity=0}, only marks, forget plot]
table {%
0.79346 1
};
\end{axis}

\end{tikzpicture}
    };
  \end{tikzpicture}
  \caption{Verification of the input at different stress levels}
  \label{fig:element-SN}
\end{figure}
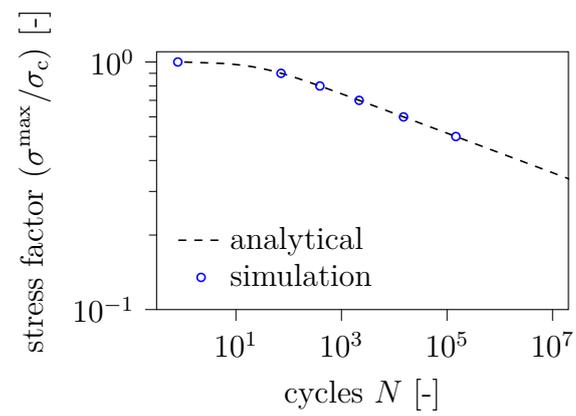

The exercise is repeated with four different stress levels. The corresponding number of cycles to failure is plotted on the underlying S-N curve that serves as input for the model in \cref{fig:element-SN}. A good match is obtained with the numerical simulations, which verifies the XFEM implementation with shifted cohesive relation and implicit fatigue damage update.

\begin{figure}
\begin{subfigure}{0.5\textwidth}
  \begin{tikzpicture}
  \node[] at (0,0)
    {
\begin{tikzpicture}

\definecolor{darkgray176}{RGB}{176,176,176}

\begin{axis}[
height=5cm,
legend cell align={left},
legend style={
  fill opacity=0.8,
  draw opacity=1,
  text opacity=1,
  at={(0.03,0.97)},
  anchor=north west,
  draw=none
},
minor xtick={},
minor ytick={},
tick align=outside,
tick pos=left,
width=7cm,
x grid style={darkgray176},
xlabel={cycles \(\displaystyle N\) [-]},
xmin=0, xmax=20000,
xtick style={color=black},
xtick={0,5000,10000,15000,20000},
y grid style={darkgray176},
ylabel={damage \(\displaystyle \mathcal{D}\) [-]},
ymin=0, ymax=0.4,
ytick style={color=black},
ytick={0,0.1,0.2,0.3,0.4}
]
\addplot [semithick, black]
table {%
0 0
1582.03308105469 0.0081632137298584
3009.13037109375 0.0163265466690063
4295.36474609375 0.0244897603988647
5453.6328125 0.0326530933380127
6495.744140625 0.0408163070678711
7432.5048828125 0.048979640007019
8273.794921875 0.0571428537368774
9028.640625 0.0653060674667358
9705.28515625 0.0734694004058838
10311.2451171875 0.0816326141357422
10853.373046875 0.0897959470748901
11337.908203125 0.0979591608047485
11770.5302734375 0.106122493743896
12156.400390625 0.114285707473755
12500.208984375 0.122448921203613
12806.212890625 0.130612254142761
13078.2705078125 0.13877546787262
13319.87890625 0.146938800811768
13534.201171875 0.155102014541626
13724.1005859375 0.163265347480774
13892.1611328125 0.171428561210632
14040.71484375 0.17959189414978
14171.8662109375 0.187755107879639
14287.5068359375 0.195918321609497
14389.3427734375 0.204081654548645
14478.904296875 0.212244868278503
14557.5673828125 0.220408201217651
14626.5625 0.22857141494751
14686.9951171875 0.236734628677368
14739.8525390625 0.244897961616516
14786.017578125 0.253061294555664
14826.27734375 0.261224508285522
14861.3359375 0.269387722015381
14891.81640625 0.277551054954529
14918.275390625 0.285714268684387
14941.2060546875 0.293877601623535
14961.046875 0.302040815353394
14978.1845703125 0.310204029083252
14992.9609375 0.3183673620224
15005.6806640625 0.326530575752258
15016.6083984375 0.334693908691406
15025.98046875 0.342857122421265
15034.001953125 0.351020336151123
15040.8544921875 0.359183669090271
15046.6962890625 0.367347002029419
15051.6669921875 0.375510215759277
15055.8876953125 0.383673429489136
15059.462890625 0.391836762428284
15062.484375 0.399999976158142
};
\addlegendentry{exact}
\addplot [blue, dashed, mark=square, mark size=1, mark options={solid,fill opacity=0}]
table {%
0 0
0 0
0 0
0 0
0 0
0 0
0 0
0 0
0 0
0 0
0 0
0 0
0 0
0 0
0.100000023841858 4.76837158203125e-07
0.100000023841858 4.76837158203125e-07
0.382843017578125 1.9073486328125e-06
0.382843017578125 1.9073486328125e-06
1.18283998966217 5.7220458984375e-06
1.18283998966217 5.7220458984375e-06
3.44558000564575 1.68085098266602e-05
3.44558000564575 1.68085098266602e-05
7.97106981277466 3.89814376831055e-05
7.97106981277466 3.89814376831055e-05
17.0219993591309 8.32080841064453e-05
17.0219993591309 8.32080841064453e-05
35.1240005493164 0.000171780586242676
35.1240005493164 0.000171780586242676
71.3277969360352 0.000349283218383789
71.3277969360352 0.000349283218383789
143.735992431641 0.000705361366271973
143.735992431641 0.000705361366271973
288.550994873047 0.00142240524291992
288.550994873047 0.00142240524291992
493.351013183594 0.00244760513305664
493.351013183594 0.00244760513305664
782.981994628906 0.00392043590545654
782.981994628906 0.00392043590545654
1192.57995605469 0.00605106353759766
1192.57995605469 0.00605106353759766
1771.83996582031 0.00916600227355957
1771.83996582031 0.00916600227355957
2351.11010742188 0.0124086141586304
2351.11010742188 0.0124086141586304
2930.3701171875 0.0157903432846069
2930.3701171875 0.0157903432846069
3509.6298828125 0.0193244218826294
3509.6298828125 0.0193244218826294
4088.88989257812 0.0230256319046021
4088.88989257812 0.0230256319046021
4668.14990234375 0.0269113779067993
4668.14990234375 0.0269113779067993
5247.419921875 0.0310020446777344
5247.419921875 0.0310020446777344
5826.68017578125 0.0353213548660278
5826.68017578125 0.0353213548660278
6405.93994140625 0.0398976802825928
6405.93994140625 0.0398975610733032
6985.2001953125 0.0447649955749512
6985.2001953125 0.0447649955749512
7564.4599609375 0.0499650239944458
7564.4599609375 0.0499650239944458
8143.72021484375 0.0555485486984253
8143.72021484375 0.0555485486984253
8722.990234375 0.0615799427032471
8722.990234375 0.0615799427032471
9302.25 0.0681413412094116
9302.25 0.0681413412094116
9881.509765625 0.0753400325775146
9881.509765625 0.0753400325775146
10460.7998046875 0.0833208560943604
10460.7998046875 0.0833208560943604
11040 0.0922852754592896
11040 0.0922852754592896
11619.2998046875 0.102527022361755
11619.2998046875 0.102527022361755
12198.599609375 0.114500045776367
12198.599609375 0.114500045776367
12777.7998046875 0.128963947296143
12777.7998046875 0.128963947296143
13357.099609375 0.147357940673828
13357.099609375 0.147357940673828
13936.2998046875 0.173064947128296
13936.2998046875 0.173064947128296
14345.900390625 0.200322031974792
14345.900390625 0.200322031974792
14635.599609375 0.230906009674072
14635.599609375 0.230906009674072
14840.400390625 0.270117998123169
14840.400390625 0.270117998123169
14956.2001953125 0.320433974266052
14956.2001953125 0.320433974266052
15002.599609375 0.378044962882996
15002.599609375 0.378042936325073
15011 0.397583961486816
15011 0.397580981254578
15011.900390625 0.40016496181488
};
\addlegendentry{simulation}
\end{axis}

\end{tikzpicture}
    };
  \end{tikzpicture}
  \caption{adaptive $\Delta N$}
  \label{fig:DvsN-adaptive}
\end{subfigure}
\begin{subfigure}{0.5\textwidth}
   \begin{tikzpicture}
  \node[] at (0,-4)
    {
\begin{tikzpicture}

\definecolor{darkgray176}{RGB}{176,176,176}

\begin{axis}[
height=5cm,
minor xtick={},
minor ytick={},
tick align=outside,
tick pos=left,
width=7cm,
x grid style={darkgray176},
xlabel={\emph{pseudo} time step},
xmin=1.5, xmax=56.5,
xtick style={color=black},
xtick={0,10,20,30,40,50,60},
y grid style={darkgray176},
ylabel={cycles \(\displaystyle N\) [-]},
ymin=-750.595, ymax=16000,
ytick style={color=black},
ytick={0,4000,8000,12000,16000}
]
\addplot [blue, mark=square, mark size=1, mark options={solid,fill opacity=0}]
table {%
4 0
4 0
5 0
5 0
6 0
6 0
7 0
7 0
8 0
8 0
9 0
9 0
10 0
10 0
11 0.1
11 0.1
12 0.382843
12 0.382843
13 1.18284
13 1.18284
14 3.44558
14 3.44558
15 7.97107
15 7.97107
16 17.022
16 17.022
17 35.124
17 35.124
18 71.3278
18 71.3278
19 143.736
19 143.736
20 288.551
20 288.551
21 493.351
21 493.351
22 782.982
22 782.982
23 1192.58
23 1192.58
24 1771.84
24 1771.84
25 2351.11
25 2351.11
26 2930.37
26 2930.37
27 3509.63
27 3509.63
28 4088.89
28 4088.89
29 4668.15
29 4668.15
30 5247.42
30 5247.42
31 5826.68
31 5826.68
32 6405.94
32 6405.94
33 6985.2
33 6985.2
34 7564.46
34 7564.46
35 8143.72
35 8143.72
36 8722.99
36 8722.99
37 9302.25
37 9302.25
38 9881.51
38 9881.51
39 10460.8
39 10460.8
40 11040
40 11040
41 11619.3
41 11619.3
42 12198.6
42 12198.6
43 12777.8
43 12777.8
44 13357.1
44 13357.1
45 13936.3
45 13936.3
46 14345.9
46 14345.9
47 14635.6
47 14635.6
48 14840.4
48 14840.4
49 14956.2
49 14956.2
50 15002.6
50 15002.6
51 15011
51 15011
52 15011.9
52 15011.9
53 15011.9
53 15011.9
54 15011.9
54 15011.9
};
\end{axis}

\end{tikzpicture}
    };
  \end{tikzpicture}
  \caption{elapsed number of cycles in each \emph{pseudo} time step}
  \label{fig:ivsN}
  \vspace{-0.3cm}
\end{subfigure}
\caption{Damage evolution and elapsed cycles per \emph{pseudo} time step with adaptive cycle increment $\Delta N$}
\label{fig:element-adaptive}  
\end{figure}
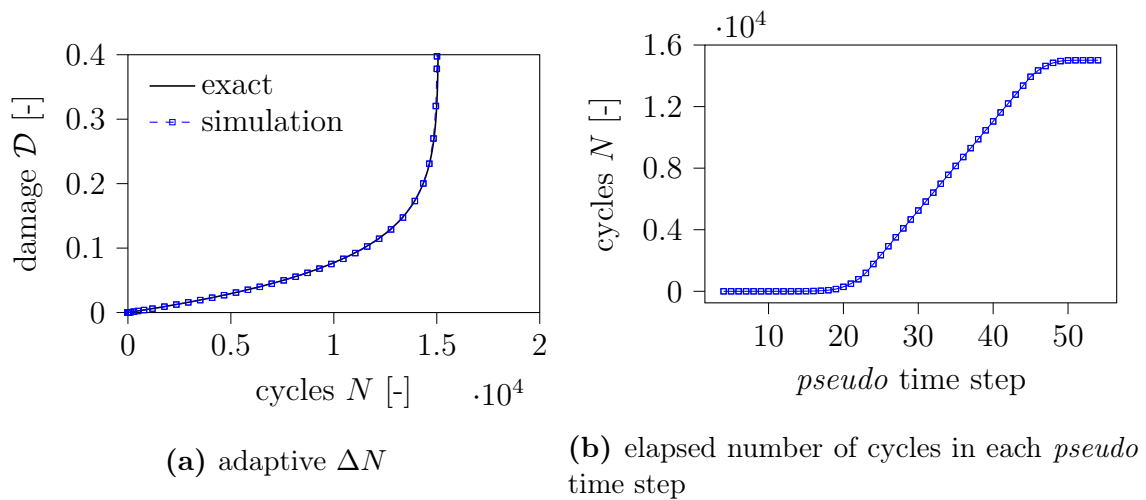

\subsection{Example B: Double cantilever beam test}

\begin{figure}
  \centering
   {\def\svgwidth{0.8\columnwidth}{\scalebox{1.0}{
\begingroup%
  \makeatletter%
  \providecommand\color[2][]{%
    \errmessage{(Inkscape) Color is used for the text in Inkscape, but the package 'color.sty' is not loaded}%
    \renewcommand\color[2][]{}%
  }%
  \providecommand\transparent[1]{%
    \errmessage{(Inkscape) Transparency is used (non-zero) for the text in Inkscape, but the package 'transparent.sty' is not loaded}%
    \renewcommand\transparent[1]{}%
  }%
  \providecommand\rotatebox[2]{#2}%
  \newcommand*\fsize{\dimexpr\f@size pt\relax}%
  \newcommand*\lineheight[1]{\fontsize{\fsize}{#1\fsize}\selectfont}%
  \ifx\svgwidth\undefined%
    \setlength{\unitlength}{295.0220111bp}%
    \ifx\svgscale\undefined%
      \relax%
    \else%
      \setlength{\unitlength}{\unitlength * \real{\svgscale}}%
    \fi%
  \else%
    \setlength{\unitlength}{\svgwidth}%
  \fi%
  \global\let\svgwidth\undefined%
  \global\let\svgscale\undefined%
  \makeatother%
  \begin{picture}(1,0.15524367)%
    \lineheight{1}%
    \setlength\tabcolsep{0pt}%
    \put(0,0){\includegraphics[width=\unitlength,page=1]{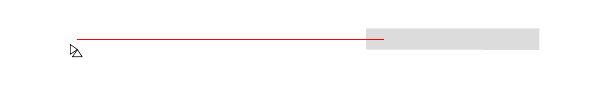}}%
    \put(-0.00166816,0.0858171){\color[rgb]{0,0,0}\makebox(0,0)[lt]{\lineheight{1.25}\smash{\begin{tabular}[t]{l}$2h$\end{tabular}}}}%
    \put(0,0){\includegraphics[width=\unitlength,page=2]{dcb-dimensions.pdf}}%
    \put(0.33098186,0.00473956){\color[rgb]{0,0,0}\makebox(0,0)[lt]{\lineheight{1.25}\smash{\begin{tabular}[t]{l}$a_0$\end{tabular}}}}%
    \put(0,0){\includegraphics[width=\unitlength,page=3]{dcb-dimensions.pdf}}%
    \put(0.13152686,0.13997508){\color[rgb]{0,0,1}\makebox(0,0)[lt]{\lineheight{1.25}\smash{\begin{tabular}[t]{l}$u^{\mathrm{max}}_{\mathrm{p}}$\end{tabular}}}}%
    \put(0,0){\includegraphics[width=\unitlength,page=4]{dcb-dimensions.pdf}}%
  \end{picture}%
\endgroup%
}}}
  \caption{Computational model of the DCB specimen. The element width is $\SI{0.05}{mm}$ in the refined zone (region in dark grey) and $\SI{0.25}{mm}$ outside this zone (region in light brown). Eight elements are used across the thickness of each arm}
  \label{fig:dcb-dimensions}
\end{figure}

The improved performance of the fatigue CZM in simulating fatigue crack growth in a DCB test is compared to the existing formulation in \cite{Davila2020}. For this purpose, the case recently studied in \cite{lecinanaRobust2023} with optimal fatigue parameters is used for comparison. The results are compared with the experiments done by \citeauthor{Renart2018} \cite{Renart2018}. The plies have height $h=\SI{1.472}{mm}$, width \SI{25}{mm} and an initial crack length $a_{0}=\SI{51.2}{mm}$. The computational model is shown in \cref{fig:dcb-dimensions}.

Each arm of the DCB specimen is modeled with 2D plane strain linear quadrilateral elements with unit thickness. The bulk material is linearly elastic orthotropic and the static interface material properties and optimal fatigue model parameters determined in \cite{lecinanaRobust2023} are given in \cref{tab:dcb-properties-ply}. Furthermore, normal dummy stiffness $K_n=\SI{2e5}{N/mm}$ is used. In order to drive the fatigue cracking process, displacement control is used with a maximum applied displacement $u^{\mathrm{max}}_{\mathrm{p}} = \SI{5}{\milli\meter}$ and \emph{global} load-ratio $R=0.1$.
 
\subsubsection*{Comparison integration schemes with constant cycle increments}
\begin{figure}
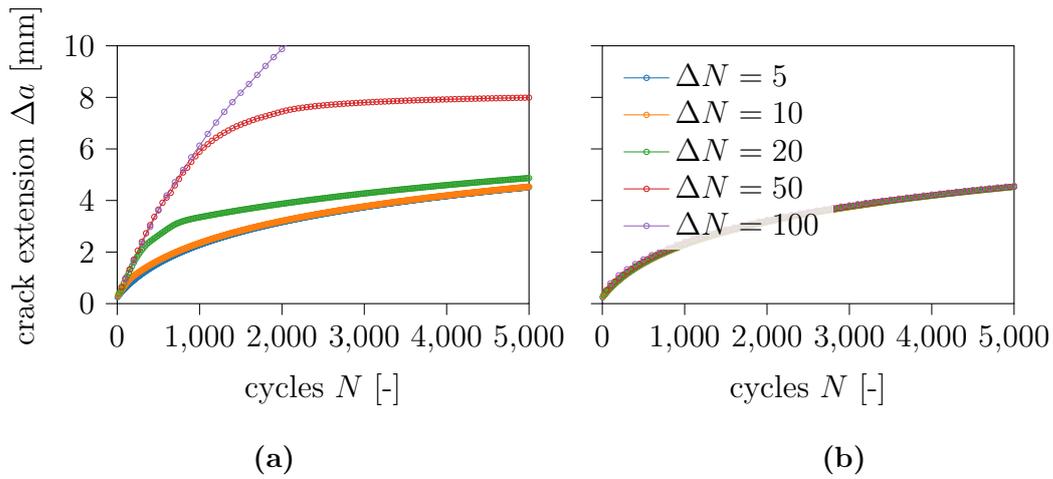

\begin{subfigure}{0.5\textwidth}
  \begin{tikzpicture}
  \node[] at (-2.5,0)
    {
      \input{figures/dcb-avsN-expl.tex}
    };
  \end{tikzpicture}
  \caption{}
  \label{fig:avsN-explicit}
\end{subfigure}%
\begin{subfigure}{0.5\textwidth}
   \begin{tikzpicture}
  \node[] at (2.5,0.0)
    {
      \input{figures/dcb-avsN-impl.tex}
    };
  \end{tikzpicture}
  \caption{}
  \label{fig:avsN-implicit}
\end{subfigure}
\caption{Crack extension with number of cycles as a result of \textbf{(a)} explicit and \textbf{(b)} implicit fatigue damage update} 
\label{fig:avsN}
\end{figure}

To compare the performance of the implicit damage update with the explicit update, constant cycle increments $\Delta N\in\{10,20,50,100\}$ are used. For the implicit scheme, the numerical time integration parameter is set to $\theta=0.5$, which reduces to the trapezoidal rule with second-order accuracy. \Cref{fig:avsN} shows the crack extension $\Delta a$ as a function of number of cycles $N$, which clearly indicates the strong influence of the step size for the explicit fatigue damage update, whereas with the implicit scheme, the global response is insensitive to the step size in the investigated range. 
 
\begin{figure}
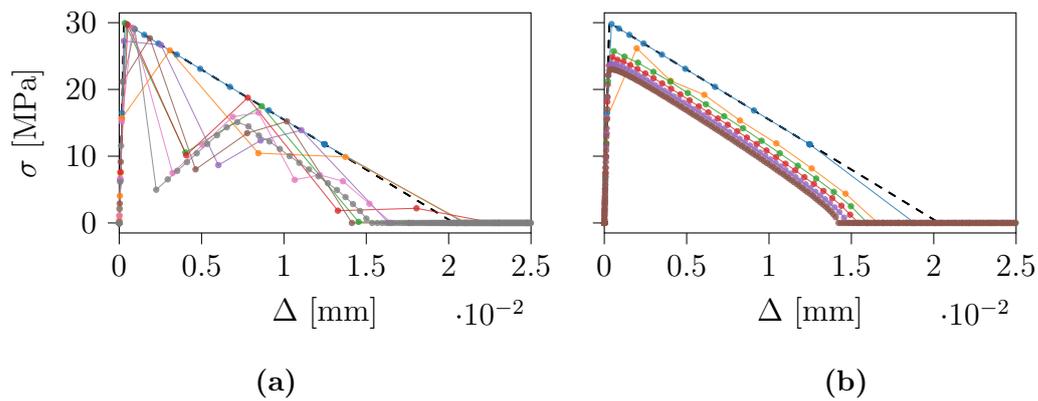

\begin{subfigure}{0.5\textwidth}
  \begin{tikzpicture}
  \node[] at (0,0)
    {
      \input{figures/dcb-tvsu-expl.tex}
    };
  \end{tikzpicture}
  \caption{}
  \label{fig:tvsu-explicit}
\end{subfigure}%
\begin{subfigure}{0.5\textwidth}
   \begin{tikzpicture}
  \node[] at (0,0)
    {
      \input{figures/dcb-tvsu-impl.tex}
    };
  \end{tikzpicture}
  \caption{}
  \label{fig:tvsu-implicit}
\end{subfigure}
\caption{Traction-opening histories of several integration points along in the interface as a result of \textbf{(a)} explicit and \textbf{(b)} implicit fatigue damage update.}
\label{fig:comparison-integration}  
\end{figure}

\Cref{fig:comparison-integration} shows the traction-opening history of several integration points along the interface for the case with $\Delta N=50$. With the explicit scheme, numerically unstable damage growth can be observed which is fully removed with the implicit scheme where damage grows during fatigue loading in a continuous fashion. The effect of numerically unstable damage growth is most profound for integration points in the cohesive zone (for which $\Dam\in(0,1)$) after the static ramp-up phase, where a sharp drop in stiffness is observed for the first fatigue cycles (see \cref{fig:comparison-integration}).
 
Under displacement control, the energy release rate (ERR) reduces as the crack extends and the compliance increases. Since the ERR is closely related to the area under the traction-opening relation, points move farther away from the softening line, which in turn decreases the rate of fatigue damage accumulation $\mathrm{d}\Dam/\mathrm{d}N$ as described by the underlying S-N curve. Consequently, larger step sizes can be used as the crack propagates.
 
\subsubsection*{Simulations with adaptive cycle jumping}
Under displacement control, the energy release rate (ERR) reduces and a sweep over several ERR values is performed in a single DCB simulation. Therefore, the crack growth rate can be computed as a function of the ERR at the maximum load level and compared to the response of the experimental Paris relation. The crack length $a$ is computed numerically and is based on the average damage along the interface following \cite{Joosten2022}:

\begin{equation}
    a = a_0 + \sum^{N_{\mathrm{ip}}}_{\mathrm{ip}=1} \Dam_{\mathrm{ip}}  J_{\mathrm{ip}} w_{\mathrm{ip}} 
\end{equation}
where $J_{\mathrm{ip}} w_{\mathrm{ip}}$ is the product of the Jacobian and the integration weight.  The crack growth rate at time step $n$ is then approximated with Euler backward differentiation:

\begin{equation}
    \dv{a}{N}^{(n)} \approx \frac{a^{(n)}-a^{(n-1)}}{\Delta N}
\end{equation}

\begin{figure}
  \centering
  \begin{tikzpicture}
    \node[] at (0,0)
    {
\begin{tikzpicture}

\definecolor{darkgray176}{RGB}{176,176,176}
\definecolor{gray}{RGB}{128,128,128}

\pgfplotsset{
  log x ticks with fixed point/.style={
      xticklabel={
        \pgfkeys{/pgf/fpu=true}
        \pgfmathparse{exp(\tick)}%
        \pgfmathprintnumber[fixed relative, precision=3]{\pgfmathresult}
        \pgfkeys{/pgf/fpu=false}
      }
  },
  log y ticks with fixed point/.style={
      yticklabel={
        \pgfkeys{/pgf/fpu=true}
        \pgfmathparse{exp(\tick)}%
        \pgfmathprintnumber[fixed relative, precision=3]{\pgfmathresult}
        \pgfkeys{/pgf/fpu=false}
      }
  }
}

\begin{axis}[
height=5cm,
legend cell align={left},
legend style={
  fill opacity=0.,
  draw opacity=1,
  text opacity=1,
  at={(0.03,0.97)},
  anchor=north west,
  draw=none
},
log basis y={10},
minor ytick={2e-08,3e-08,4e-08,5e-08,6e-08,7e-08,8e-08,9e-08,2e-07,3e-07,4e-07,5e-07,6e-07,7e-07,8e-07,9e-07,2e-06,3e-06,4e-06,5e-06,6e-06,7e-06,8e-06,9e-06,2e-05,3e-05,4e-05,5e-05,6e-05,7e-05,8e-05,9e-05,0.0002,0.0003,0.0004,0.0005,0.0006,0.0007,0.0008,0.0009,0.002,0.003,0.004,0.005,0.006,0.007,0.008,0.009,0.02,0.03,0.04,0.05,0.06,0.07,0.08,0.09,0.2,0.3,0.4,0.5,0.6,0.7,0.8,0.9,2,3,4,5,6,7,8,9},
tick align=outside,
tick pos=left,
unbounded coords=jump,
width=7cm,
x grid style={darkgray176},
xlabel={\(\displaystyle \frac{G^{\mathrm{max}}_I(1-R)}{G_{Ic}}\) [-]},
xmin=0.2, xmax=1.0,
xmode=log,
log x ticks with fixed point,
xtick style={color=black},
xtick={0.2,0.4,0.6,0.8,1.0},
y grid style={darkgray176},
ylabel={\(\displaystyle \frac{\mathrm{d}a}{\mathrm{d}N}\) [mm]},
ymin=1e-07, ymax=0.01,
ymode=log,
ytick style={color=black},
ytick={1e-08,1e-07,1e-06,1e-05,0.0001,0.001,0.01,0.1,1}
]
\addplot [line width=0.12pt, gray, opacity=0.5, mark=*, mark size=1.5, mark options={solid}, only marks]
table {%
0.292134696627527 1.1262457517e-06
0.293694928649918 1.3239350697e-06
0.297225288160641 8.0410605306124e-07
0.297653487071974 1.3217066868e-06
0.299157047079265 1.96363558e-06
0.298726685173758 7.84042028777989e-07
0.300235666309361 1.4500168255e-06
0.30066820213514 1.3217066868e-06
0.302186990605054 1.5123872971e-06
0.305247622218785 2.1542636189e-06
0.305247622218785 9.27895911232042e-07
0.30774796623925 3.0599571254e-06
0.309896790576952 1.8826557427e-06
0.311126087759434 1.2512904042e-06
0.312135502544743 2.7043614431e-06
0.313260915964643 1.4402790546e-06
0.315070023557352 9.27895911232042e-07
0.317934038832909 1.3328862353e-06
0.318490307656378 1.4257948771e-06
0.319485284499875 1.3863207715e-06
0.326929678359035 1.6314921321e-06
0.329568084375911 1.6178080343e-06
0.32968671872901 1.2996281077e-06
0.330637332215367 1.2149408331e-06
0.33543167281463 4.4827496688e-06
0.340540568694773 2.5069332625e-06
0.351245722256451 4.226087805e-06
0.353443762502525 4.214239216e-06
0.356082487117281 8.5026083051e-06
0.365430286606382 9.7292666533e-06
0.366835862554483 8.1291073994e-06
0.368158499676032 1.60820173442e-05
0.368865854960341 8.2209155931e-06
0.371530562789267 8.1519628777e-06
0.375293524866699 1.42731200335e-05
0.374483980496666 1.30650164708e-05
0.377460875885387 1.25262184785e-05
0.386905454360377 1.1070554551e-05
0.387369896155724 1.01477576248e-05
0.390730375862787 1.10395162817e-05
0.391293283287672 1.61272329109e-05
0.393836419150931 2.36259349483e-05
0.394972000871987 1.42131362793e-05
0.395825835418401 1.36269901149e-05
0.398398429737422 2.35596954363e-05
0.399834866591409 2.47809326325e-05
0.399738943212666 2.34936416385e-05
0.413295500114798 3.52982742672e-05
0.424145243094572 2.52019205894e-05
0.428441100891858 4.58301569951e-05
0.428955402129682 4.18921569959e-05
0.431536183671755 5.20021479901e-05
0.431225671891435 4.76674579707e-05
0.440321881945406 3.35587285392e-05
0.448640113989097 4.00519265609e-05
0.458103661689267 7.28351664997e-05
0.463076666268029 7.85711446449e-05
0.466421958333975 0.0001100481926894
0.475347292897593 0.0001465395231125
0.477863094260715 0.000111291050643
0.480392210639181 0.0001187148033189
0.483514427941358 0.000136990612349
0.490172574778699 0.000158079930616
0.49205795430738 0.0001648795205637
0.496922406407114 0.0002052459557384
0.504127930513417 0.0002251710550917
0.508379306720783 0.0002569337553788
0.508135407975285 0.0001690988692503
0.516617918704591 0.0001934945055301
0.518107526665468 0.0002087327524962
0.520349995810734 0.0001934945055301
0.526150192871766 0.0002356490756058
0.52789522699811 0.0002429039211373
0.530714608082034 0.0002894148688597
0.535549866303347 0.000264249688595
0.539030570240087 0.0002802982944693
0.542455797832118 0.0003159104531172
0.546059991650779 0.0003860328809556
0.546453191942927 0.0003075094097664
0.555175592857314 0.0003459954214002
0.557578499731105 0.0003992604403362
0.562010873213404 0.0003992604403362
0.567703071419467 0.0004060432293266
0.573452921749291 0.0004984062798824
0.578150188753344 0.0006134992983282
0.580095523023988 0.0005687116274443
0.586139626551579 0.0005971844402724
0.58723761569556 0.0006730572780622
0.591054254274638 0.0006787502783407
0.59361245090405 0.0006674120276752
0.596181719901096 0.0006787502783407
0.600920962980225 0.0006674120276752
0.614126935956653 0.0005951758286114
0.615365935263757 0.0005687116274443
0.631520387064625 0.0010255381432974
0.642987603671043 0.0009347895815541
0.645615652639625 0.0009722646836
0.662882236070547 0.0015495072590304
0.680610603385051 0.0012983019280289
0.702006118911642 0.0018493177079641
0.707586594678509 0.0017288111757213
0.710137818242954 0.0016298273083999
0.718363711246135 0.0013656005317289
0.742725957390249 0.0017336718331497
0.745225048952813 0.0018338066038623
0.326073640776378 9.73274386158155e-07
0.295861768163255 7.03167554794648e-07
0.323816884063852 5.36304899694288e-07
0.319350119600823 2.7993604566e-06
0.333710370494769 3.3838551534e-06
0.336814909156308 3.6703364977e-06
0.345502121899261 4.2027098876e-06
0.343110900606636 4.2027098876e-06
0.343906129841572 4.8123027439e-06
0.344703202181857 5.2197182204e-06
0.354413397363874 3.1197345819e-06
0.356883388438682 5.5103155628e-06
0.359370593467464 6.4828310849e-06
0.361875132418349 7.4231499801e-06
0.363554514630817 7.8364189662e-06
0.347910008180553 1.38412868955e-05
0.381665552540782 1.58489319246e-05
0.403471245973975 2.9552092352e-05
0.40722477655826 2.9552092352e-05
0.408168602729148 3.20540088826e-05
0.411965833260881 3.11973458191e-05
0.418696187436677 3.2934195473e-05
0.41579838978196 2.79936045664e-05
0.42357079841341 3.57224450187e-05
0.420639263560506 4.09038988609e-05
0.44467164155769 6.84374970259e-05
0.455085967388886 5.97682566407e-05
0.454033652136921 4.94444610115e-05
0.449848668074815 9.21946844785e-05
0.486687368546094 0.0001241988707283
0.473353139715142 0.0001628413544013
0.50155699120561 0.0001673128942025
0.499240128518799 0.0001719072201858
0.514493272132819 0.0002078007251824
0.521689512633066 0.0002444754724726
0.522898634089924 0.0002193696137571
0.56050506178081 0.0002580861540418
0.65303432354387 0.0015425359490188
0.641051321807307 0.0011764899870201
0.636614603555204 0.0010846612314544
0.619172678102548 0.0007626985859023
0.616312508471295 0.0007031675547946
0.626381308187407 0.0008973072494285
0.602208625387646 0.0005363048996942
0.593901695701174 0.0005219718220435
0.659109562685369 0.0011144454707535
0.676109478176951 0.0016731289420254
0.683980986881083 0.0015848931924611
0.687155190909302 0.0017190722018585
0.698381359283385 0.0015848931924611
0.739993091599755 0.0021350682617126
0.716394182584481 0.0026517306703257
0.72809803321721 0.0029552092352028
0.719718809063532 0.0032934195473009
};
\addlegendentry{experiments}
\addplot [ultra thick, blue, opacity=1.0]
table {%
nan nan
nan nan
nan nan
0.721075053905155 0.00706109833789048
0.700198574048501 0.00350795809126614
0.682407815192606 0.00263918202772884
0.661636572772382 0.00208499854875199
0.641794940850536 0.00155157566325156
0.623894396589774 0.0012806941896143
0.602900465806318 0.000923723880045108
0.587002590611647 0.000708362365905782
0.568477027581458 0.000563370135567634
0.551148637790729 0.000411668950494187
0.537976698438714 0.000338400711373844
0.524145236723751 0.000273530982824784
0.508678247825198 0.000203985850221553
0.497893497649571 0.000167146652516256
0.484640716910056 0.000139822388810591
0.470662399649566 0.000104789983998597
0.459700059240094 8.77675256079607e-05
0.443824099257063 6.34639046429844e-05
0.431883319487498 4.95262700560923e-05
0.421255915527154 3.98884331102552e-05
0.408710077240198 3.07076127026794e-05
0.398140969834279 2.44874483905952e-05
0.388726081376063 2.02354770696884e-05
0.376226681409994 1.50682030703375e-05
0.366950343115843 1.16544161699614e-05
0.35812348155419 9.99252186766162e-06
0.347404348207445 7.40401094391637e-06
0.338967100123874 6.12849933825563e-06
0.332263625293375 4.87690410090938e-06
0.323059082549176 3.93994116228797e-06
0.314959518697047 3.07621424441536e-06
0.308500389655879 2.68820937828558e-06
0.298828257651775 2.00123450360372e-06
0.291350661487396 1.57779099248446e-06
0.285425359316067 1.29751897130792e-06
0.278096670143227 1.07366613342228e-06
0.270842261771051 8.61998661472679e-07
0.265151270506701 6.8077280386139e-07
0.259827633753986 5.84711178128909e-07
0.252467781069291 4.69756234992802e-07
0.246863394259857 3.72350640704042e-07
};
\addlegendentry{simulation}
\end{axis}

\end{tikzpicture}
    };
  \end{tikzpicture}
  \caption{Paris relation with simulation and with experimental results from \cite{Renart2018}}
  \label{fig:paris}
\end{figure}
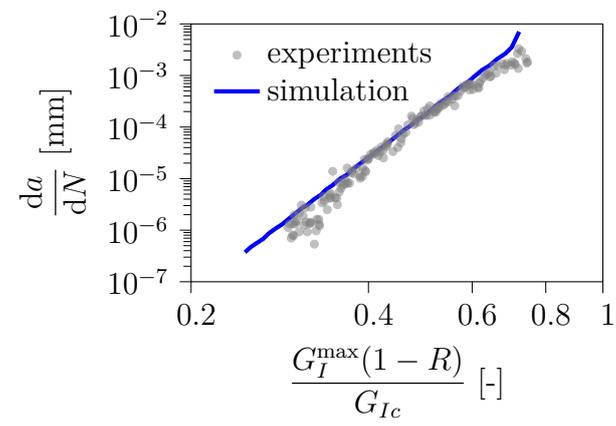

The ERR at maximum load $G^{\mathrm{max}}_{I}$ is computed according to the ASTM D5528 standard \cite{ASTM2002} 
\begin{equation}
    G^{\mathrm{max}}_I = \frac{3F^{\mathrm{max}} u^{\mathrm{max}}_\mathrm{p}}{2b(a+\Delta_{\mathrm{cor}})}
\end{equation}
where $F^{\mathrm{max}}$ is the reaction force at the node where $u^{\mathrm{max}}_{\mathrm{p}}$ is applied (see \cref{fig:dcb-dimensions}), $b$ is the specimen width and $\Delta_{\mathrm{cor}}$ takes into account the effect of finite rotations (here $\Delta_{\mathrm{cor}} =\SI{6.2}{mm}$). The computed crack growth rate vs ERR is shown in \cref{fig:paris}, from which it can be observed that a good match is obtained with the experimental results.

\begin{figure}
  \centering
  {\def\svgwidth{1.3\columnwidth}{\scalebox{0.75}{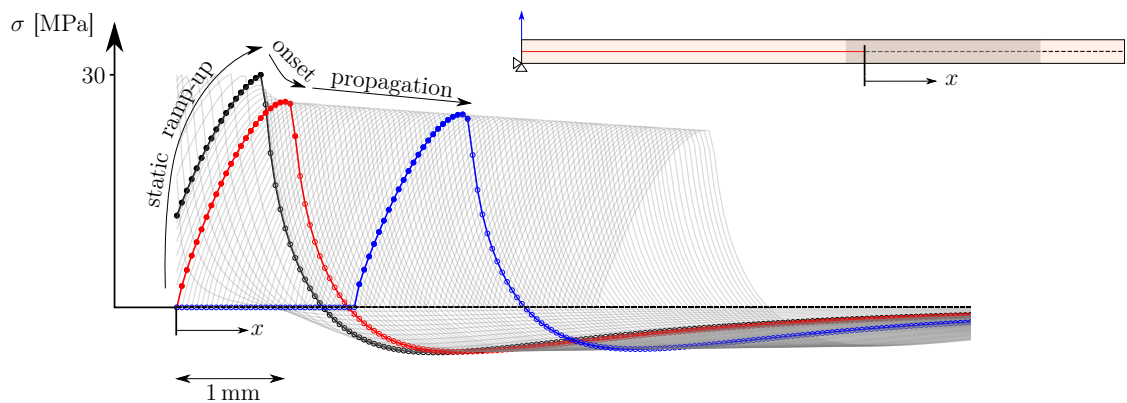}}}
  \caption{Traction profile evolution along the interface at different time steps shown in gray. The stages of the static ramp-up, fatigue crack onset and propagation are indicated with black, red and blue colors respectively. }
  \label{fig:trac-profile}
\end{figure}

The complete evolution of the traction is shown in \cref{fig:trac-profile}, from which three phases can be distinguished. First, the maximum load is applied quasi-statically. When the first load cycles are applied (shown in black), the fracture process continuous to develop into a complete cohesive zone during the \emph{onset phase} (shown in red). When the process zone has fully developed, propagation of the crack takes place (shown in blue).  

\begin{table}
  \centering
  \caption{Ply material properties for the DCB specimen example. Taken from \cite{lecinanaRobust2023}} 
  \label{tab:dcb-properties-ply}
  \begin{tabular}{lr lr lr}
  \toprule
  \multicolumn{2}{c}{\textbf{elastic constants ply}}     & \multicolumn{2}{c}{\textbf{fracture properties interface}}    & \multicolumn{2}{c}{\textbf{fatigue parameters}}                \\ 
  \midrule
  \multicolumn{1}{l}{$E_1$} & \SI{154}{GPa} & \multicolumn{1}{l}{$f_{n}$}   & 30 MPa  & \multicolumn{1}{l}{$\eta$}     & 0.8757 \\ 
  \multicolumn{1}{l}{$E_2=E_3$} & 8.5 GPa  &  \multicolumn{1}{l}{$G_{Ic}$}   & \SI{0.305}{\newton\per\milli\meter} & \multicolumn{1}{l}{$\epsilon$} & 0.2628   \\ 
  \multicolumn{1}{l}{$G_{12} = G_{13}$} & 4.2 GPa   & &  & \multicolumn{1}{l}{$p$} & $\beta + 0.915$ \\ 
  \multicolumn{1}{l}{$\nu_{12}=\nu_{13}$}  & 0.35      &   &    & \multicolumn{1}{l}{}           &           \\
  \multicolumn{1}{l}{$\nu_{23}$}        &    0.4       &     &   & \multicolumn{1}{l}{}           &           \\ \bottomrule
  \end{tabular}
\end{table}

\subsection{Example C: Open-hole $[\pm45]$-laminate}
A $[\pm45]$-laminate is simulated in order to demonstrate the capabilities of the developed fatigue failure framework to deal with progressive laminate failure. This case was previously studied in the context of static loading in \cite{Meer2010} for the purpose of simulating the interaction between matrix cracking and delamination without fiber fracture. 

\subparagraph*{Modeling preliminaries}

The laminate has dimensions $\SI{38}{mm}\times\SI{16}{mm}\times\SI{1}{mm}$ and contains a hole in the middle with a diameter of $\SI{6.4}{mm}$. The plies have a thickness $t=\SI{0.5}{\milli\meter}$ and are modeled with plane stress XFEM elements, whereas delamination between the plies is modeled with zero-thickness interface elements. The constitutive relation of the matrix cohesive segments and interface integration points is described with the improved fatigue CZM presented in \cref{sec:fczm-formulation}. 
The material properties and the fatigue model parameters are given in \cref{tab:open-hole-properties-ply}. The dummy stiffness of matrix cracks in normal direction is $K_n = \SI{1.e5}{\newton\per\milli\meter}$, whereas $K_{sh}$ is adapted according to \cref{eq:constrain-dummies}. The shear stiffness of the interface is related to the shear modulus $G_{12}$ and thickness of the ply ($K_d = G_{12}/\frac{1}{2}t=\SI{22000}{N/mm}$) according to \cite{Meer2010}, taking into account the effect of out-of-plane ply shear deformation. The computational domain is discretized using linear (constant strain) triangular elements with single-point Gauss integration. Upon crack insertion, a two-point Gauss scheme is employed in the XFEM crack segment. Three-point Newton-Cotes is used for the interface, following \cite{Meer2010}. Furthermore, displacement control is used. The maximum applied displacement is $u^{\mathrm{max}}_{\mathrm{p}} = \SI{0.1}{\milli\meter}$ with \emph{global} load ratio $R=0.1$.

Crack segments are inserted when the stress in the bulk integration point reaches the stress endurance limit (see \cref{sec:fatigue-xfem-insertion-criterion}). The shift is applied in order to ensure traction continuity before and after insertion of the crack, which improves the convergence characteristics. At the end of a converged \emph{pseudo} time step, a maximum of 100 segments can be inserted at a time. After insertion, the \emph{global} Newton-Raphson loop is re-entered to equilibrate the solution with new crack segments \cite{VanderMeer2012a}. With the XFEM representation, cracks can initiate at arbitrary locations within a predefined crack-spacing width $l_\mathrm{c}$.  

\Cref{eq:adaptive-step} is used to adapt the cycle increment $\Delta N$ throughout the simulations. The used stepping parameters are $C=2$, $\xi=2$, $n_{\mathrm{iter}}^{\mathrm{opt}}=4$, $n_{\mathrm{iter}}^{\mathrm{max}}=10$ and $c_{\mathrm{red}}=0.6$.

\begin{table}
  \centering
  \caption{Ply material properties for open-hole numerical example} 
  \label{tab:open-hole-properties-ply}
  \begin{tabular}{lr lr lr}
  \toprule
  \multicolumn{2}{c}{\textbf{elastic constants}}                       & \multicolumn{2}{c}{\textbf{fracture properties}}             & \multicolumn{2}{c}{\textbf{fatigue parameters}}                \\ 
  \midrule
  \multicolumn{1}{l}{$E_1$}                & \SI{122.7}{GPa} & \multicolumn{1}{l}{$f_{2t}$}   & 80 MPa  & \multicolumn{1}{l}{$\eta$}     & 0.95      \\ 
  \multicolumn{1}{l}{$E_2$}             & 10.1 GPa  & \multicolumn{1}{l}{$f_{12}$}   & 100 MPa & \multicolumn{1}{l}{$\epsilon$} & 0.25       \\ 
  \multicolumn{1}{l}{$G_{12}$} & 5.5 GPa   & \multicolumn{1}{l}{$G_{Ic}$}   & \SI{0.969}{\newton\per\milli\meter}   & \multicolumn{1}{l}{$p$}        & $\beta$ \\ 
  \multicolumn{1}{l}{$\nu_{12}$}  & 0.25      & \multicolumn{1}{l}{$G_{IIc}$}  & \SI{1.719}{\newton\per\milli\meter}   & \multicolumn{1}{l}{}           &           \\
  \multicolumn{1}{l}{}         &          & \multicolumn{1}{l}{$\eta$}     & 2.284   & \multicolumn{1}{l}{}           &           \\ \bottomrule
  \end{tabular}
  \end{table}

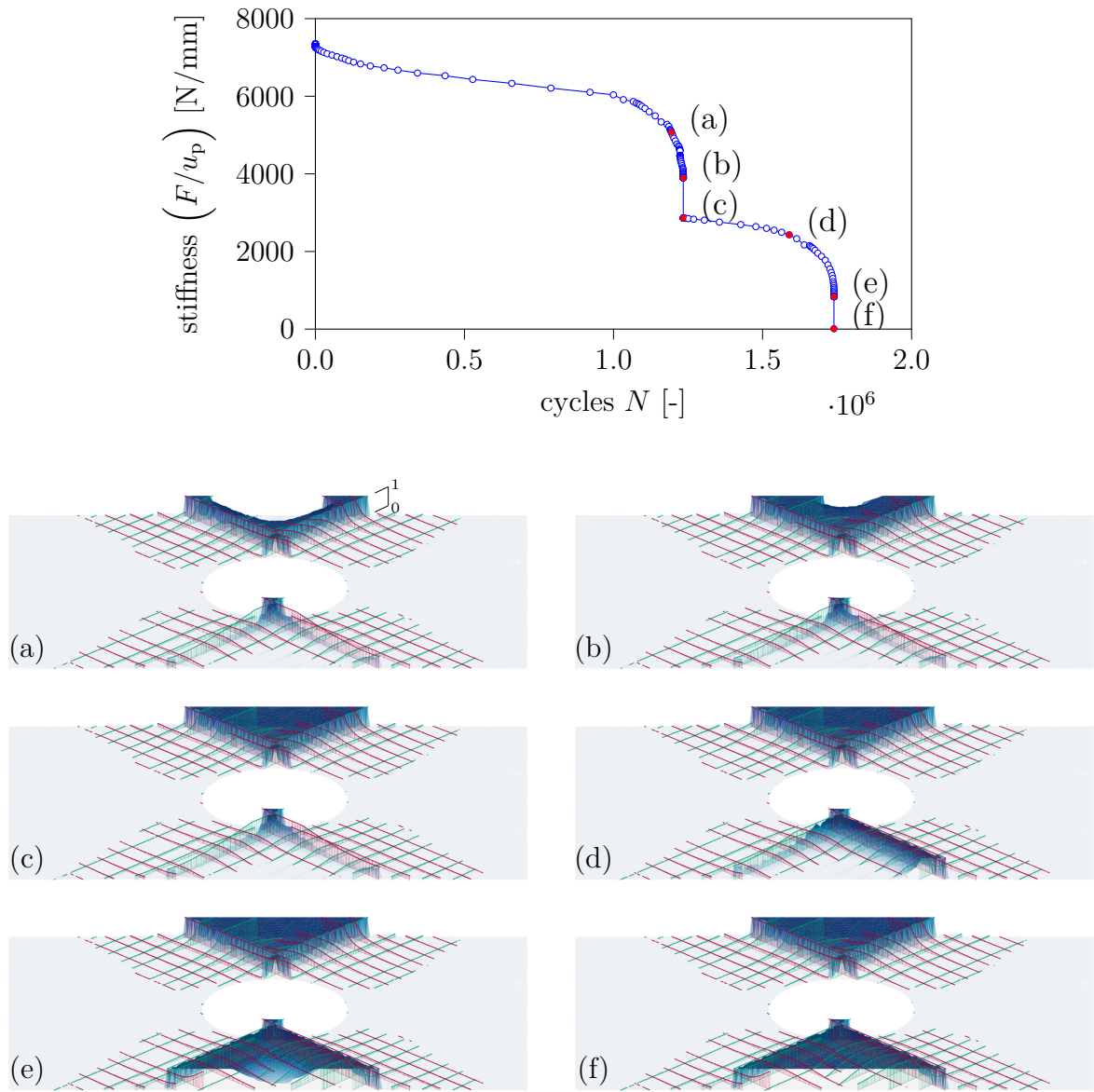
\begin{figure}
  \centering 
  \begin{tikzpicture}
    \node[] at (0,4)
    {
\begin{tikzpicture}

\definecolor{darkgray176}{RGB}{176,176,176}

\begin{axis}[
height=6cm,
minor xtick={},
minor ytick={},
tick align=outside,
tick pos=left,
width=10cm,
x grid style={darkgray176},
xlabel={cycles \(\displaystyle N\) [-]},
xmin=0, xmax=2000000,
xtick style={color=black},
xtick={0,500000,1000000,1500000,2000000},
xticklabels={
  \(\displaystyle {0.0}\),
  \(\displaystyle {0.5}\),
  \(\displaystyle {1.0}\),
  \(\displaystyle {1.5}\),
  \(\displaystyle {2.0}\)
},
y grid style={darkgray176},
ylabel={stiffness \(\displaystyle \Big(F/u_{\mathrm{p}}\Big)\) [N/mm]},
ymin=0, ymax=8000,
ytick style={color=black},
ytick={0,2000,4000,6000,8000},
yticklabels={
  \(\displaystyle {0}\),
  \(\displaystyle {2000}\),
  \(\displaystyle {4000}\),
  \(\displaystyle {6000}\),
  \(\displaystyle {8000}\)
}
]
\addplot [blue, mark=*, mark size=1.25, mark options={solid,fill=white}]
table {%
0 7349.3011666263
0 7349.3011826507
0 7349.26420129992
0 7348.9296617617
0 7347.63789206157
0 7345.96521051017
0 7342.0415237132
0 7330.98559187964
0 7280.55055
0 7280.55055
0 7280.55055
0 7280.43059
1 7280.22008
3 7279.86246
7 7279.27512
15 7278.59574
26.31 7277.79188
42.31 7276.83136
64.94 7275.67843
96.94 7274.30156
142.19 7272.66752
206.19 7270.73759
296.7 7268.47174
424.7 7265.81492
605.7 7262.69519
861.7 7259.01358
1223.7 7254.64708
1735.7 7249.4348
2459.8 7243.178
3483.8 7235.5936
4931.8 7226.29466
6979.8 7214.79834
9875.8 7200.5141
13971.8 7182.72831
19764.8 7160.48617
27956.8 7132.36152
39546.8 7095.82017
55926.8 7060.1074
72306.8 7019.25178
88686.8 6983.03153
100276.8 6952.53765
111866.8 6917.21468
128246.8 6877.8032
151416.8 6832.89902
184186.8 6779.32877
230526.8 6732.56285
276866.8 6673.42101
342406.8 6598.19939
435086.8 6528.44406
527766.8 6432.5425
658866.8 6331.70318
789966.8 6208.62226
921066.8 6102.40499
999706.8 6036.25903
1033076.8 5911.03634
1066446.8 5863.51296
1076456.8 5821.19175
1083534.8 5800.4274
1088539.8 5774.01694
1095617.8 5737.77454
1105627.8 5686.04559
1119787.8 5600.24783
1139807.8 5493.37191
1159827.8 5338.15377
1179847.8 5274.24917
1185853.8 5221.64729
1189456.8 5142.19244
1191258.8 5130.03346
1191895.8 5116.05401
1192796.7 5090.27522
1194598.7 5066.20625
1196400.7 5032.63669
1198948.7 4986.32749
1202551.7 4919.63485
1207647.7 4849.66278
1212743.7 4767.50972
1217839.7 4723.8992
1220001.7 4691.38973
1221298.7 4638.30553
1222595.7 4623.50379
1222829.2 4610.62371
1222969.3 4598.06408
1223053.36 4590.66026
1223083.62 4471.91957
1223109.3 4470.18245
1223111.57 4468.02083
1223116.109 4465.53096
1223125.188 4463.47992
1223138.028 4460.68692
1223163.708 4458.0355
1223200.018 4455.09909
1223251.378 4451.27018
1223324.008 4446.78882
1223426.708 4438.85209
1223632.108 4431.26107
1223837.508 4421.3676
1224128.008 4403.01438
1224709.008 4379.14393
1225530.708 4348.2706
1226692.708 4307.53287
1228335.708 4261.32459
1229978.708 4208.88995
1231621.708 4159.72347
1232783.708 4131.95234
1233276.708 4109.10821
1233572.508 4064.37291
1233868.308 4036.39645
1233957.048 4014.20636
1233994.698 3987.58695
1234017.288 3970.30027
1234025.421 3938.79271
1234032.322 3923.82455
1234034.079 3912.62518
1234034.9733 3905.07626
1234035.4286 3901.12358
1234035.6253 3899.16087
1234035.68539 3898.1234
1234035.69837 3897.94928
1234035.699379 3897.86586
1234035.699815 3897.78848
1234035.6999482 3897.78587
1234035.7000612 2862.15564
1234044.4420612 2862.15373
1234048.8130612 2862.14832
1234061.1730612 2862.13302
1234096.1430612 2862.08976
1234195.0530612 2861.96741
1234474.8530612 2861.62158
1235266.1530612 2860.64636
1237504.1530612 2858.70576
1241980.1530612 2854.84732
1250932.1530612 2847.15193
1268832.1530612 2831.44474
1304642.1530612 2806.97841
1355282.1530612 2755.9897
1426902.1530612 2693.46996
1477542.1530612 2640.24576
1513352.1530612 2596.57769
1538672.1530612 2549.21784
1563992.1530612 2494.63371
1589312.1530612 2430.30568
1614632.1530612 2330.3118
1639952.1530612 2168.99971
1657852.1530612 2149.44875
1661017.1530612 2125.84642
1665493.1530612 2103.85401
1669969.1530612 2073.83379
1676299.1530612 2028.80602
1685251.1530612 1955.81812
1697911.1530612 1873.38691
1710571.1530612 1774.52861
1719523.1530612 1657.06821
1725853.1530612 1549.91889
1730329.1530612 1459.90809
1733494.1530612 1381.43724
1735732.1530612 1304.39161
1737315.1530612 1220.96966
1738434.1530612 1167.01826
1738908.9530612 1117.31541
1739193.8530612 1065.84802
1739364.7530612 1027.56463
1739437.2630612 981.747766
1739480.7730612 941.821872
1739499.2330612 908.78849
1739507.0640612 865.644039
1739511.7630612 846.186383
1739512.9590612 837.86252
1739513.3244612 835.964855
1739513.3718112 835.327096
1739513.3820412 834.741718
1739513.3872482 834.671742
1739513.3875918 834.62796
1739513.38766601 9.29031514
};
\addplot [semithick, red, mark=*, mark size=1, mark options={solid}, only marks]
table {%
1194598.7 5066.20625
};
\addplot [semithick, red, mark=*, mark size=1, mark options={solid}, only marks]
table {%
1234035.6999482 3897.78587
};
\addplot [semithick, red, mark=*, mark size=1, mark options={solid}, only marks]
table {%
1234035.7000612 2862.15564
};
\addplot [semithick, red, mark=*, mark size=1, mark options={solid}, only marks]
table {%
1589312.1530612 2430.30568
};
\addplot [semithick, red, mark=*, mark size=1, mark options={solid}, only marks]
table {%
1739513.3875918 834.62796
};
\addplot [semithick, red, mark=*, mark size=1, mark options={solid}, only marks]
table {%
1739513.38766601 9.29031514
};
\draw (axis cs:1214598.7,5166.20625) node[
  scale=1.1,
  anchor=base west,
  text=black,
  rotate=0.0
]{(a)};
\draw (axis cs:1254035.6999482,3997.78587) node[
  scale=1.1,
  anchor=base west,
  text=black,
  rotate=0.0
]{(b)};
\draw (axis cs:1254035.7000612,2962.15564) node[
  scale=1.1,
  anchor=base west,
  text=black,
  rotate=0.0
]{(c)};
\draw (axis cs:1609312.1530612,2530.30568) node[
  scale=1.1,
  anchor=base west,
  text=black,
  rotate=0.0
]{(d)};
\draw (axis cs:1759513.3875918,934.62796) node[
  scale=1.1,
  anchor=base west,
  text=black,
  rotate=0.0
]{(e)};
\draw (axis cs:1759513.38766601,109.29031514) node[
  scale=1.1,
  anchor=base west,
  text=black,
  rotate=0.0
]{(f)};
\end{axis}

\end{tikzpicture}
    };
    \node at (-4,-1.7){\includegraphics[clip, width=0.5\columnwidth, trim = 0 3.5cm 0 5cm]{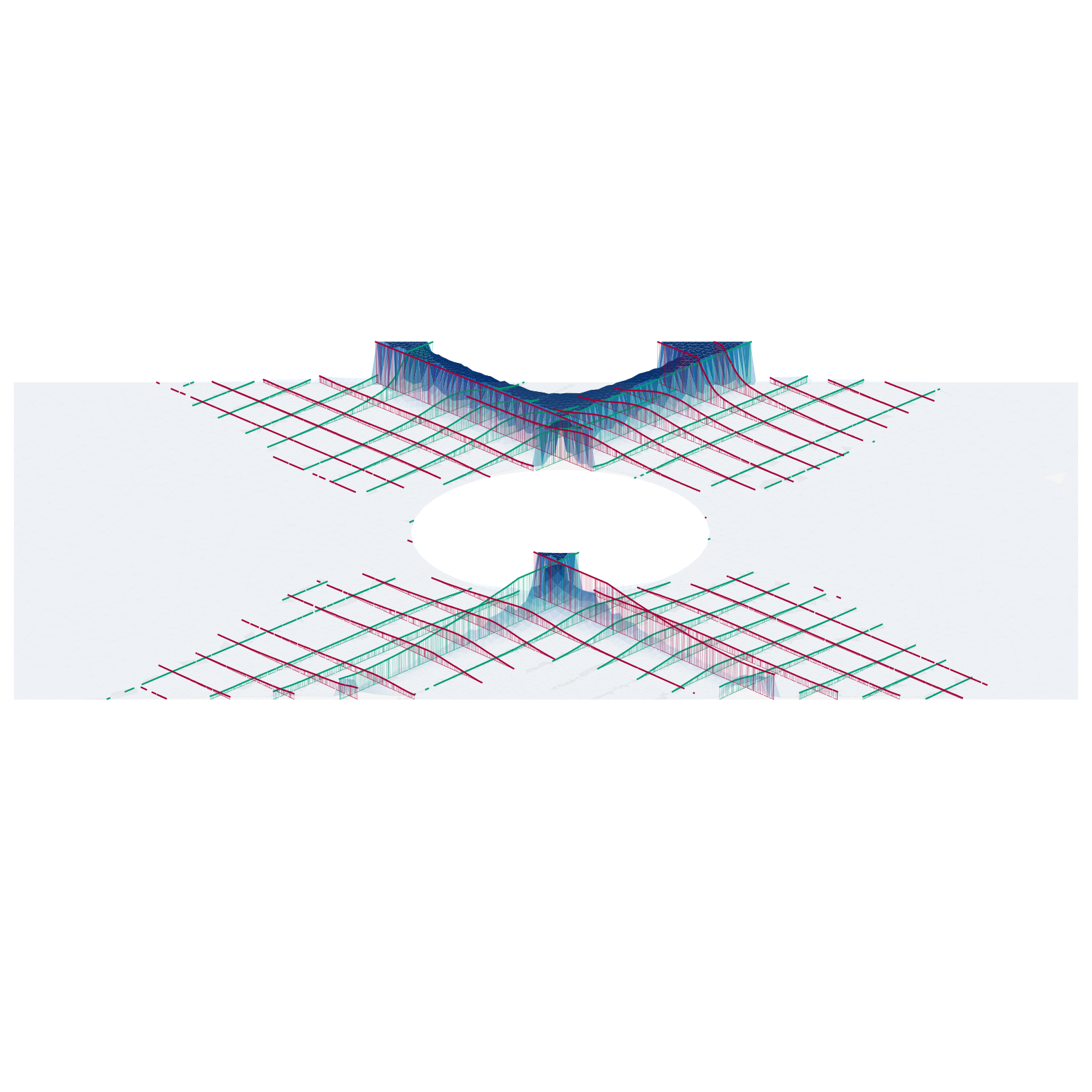}};
    \node[] at (-7.4,-2.2){(a)};
    \node at (4,-1.7) {\includegraphics[clip, width=0.5\columnwidth, trim = 0 3.5cm 0 5cm]{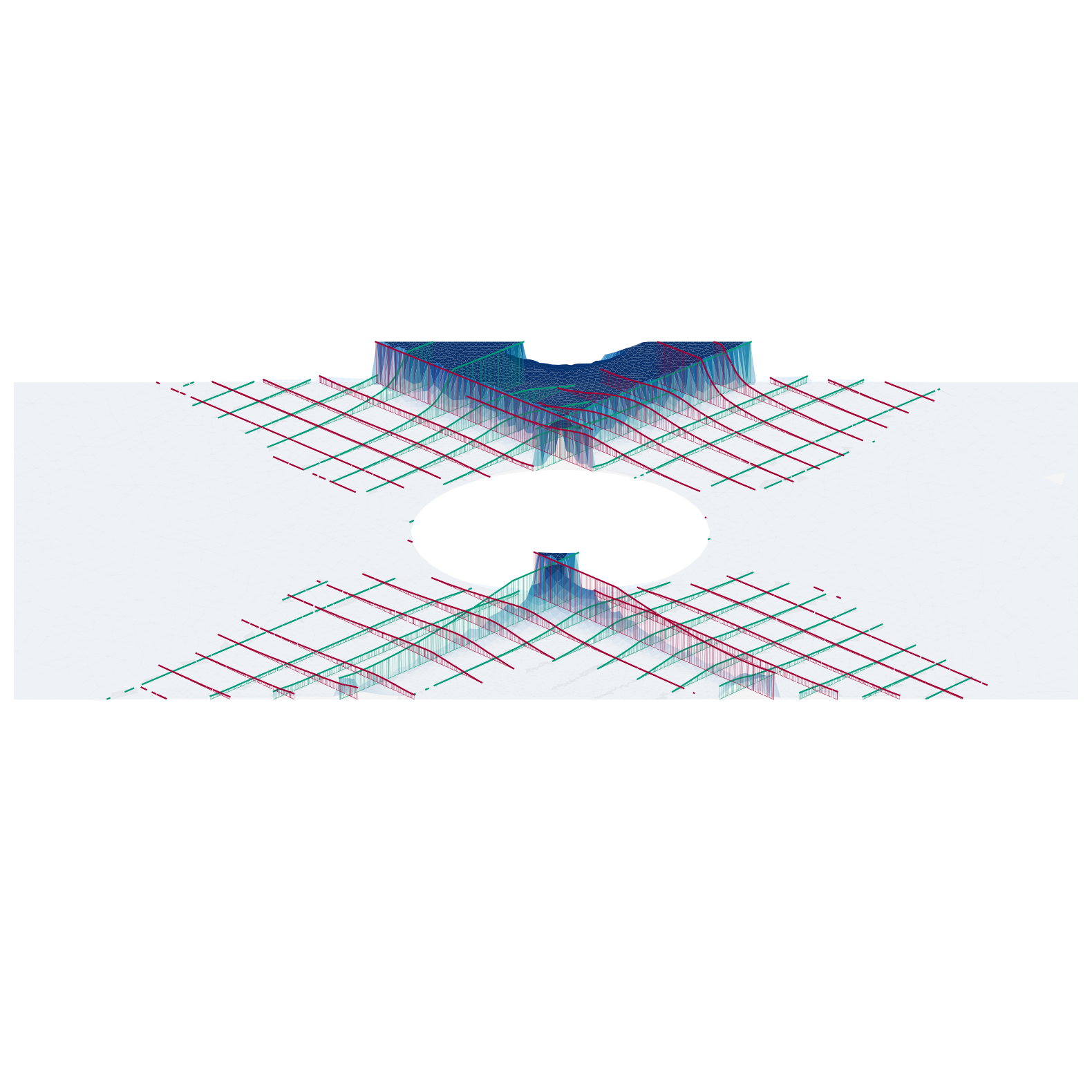}};
    \node[] at (0.6,-2.2){(b)};
    \node at (-4,-4.7){\includegraphics[clip, width=0.5\columnwidth, trim = 0 3.5cm 0 5cm]{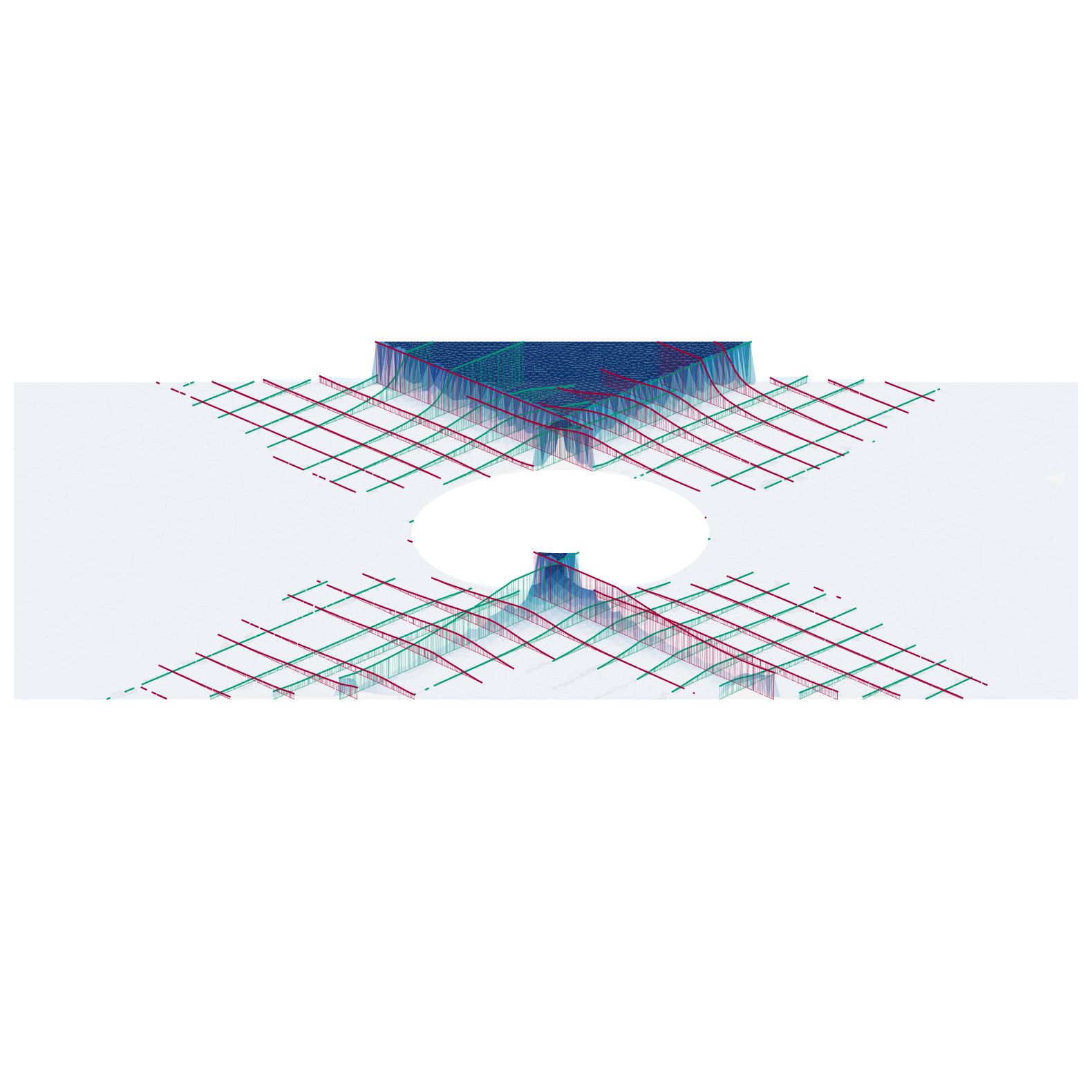}};
    \node[] at (-7.4,-5.2){(c)};
    \node at (4,-4.7) {\includegraphics[clip, width=0.5\columnwidth, trim = 0 3.5cm 0 5cm]{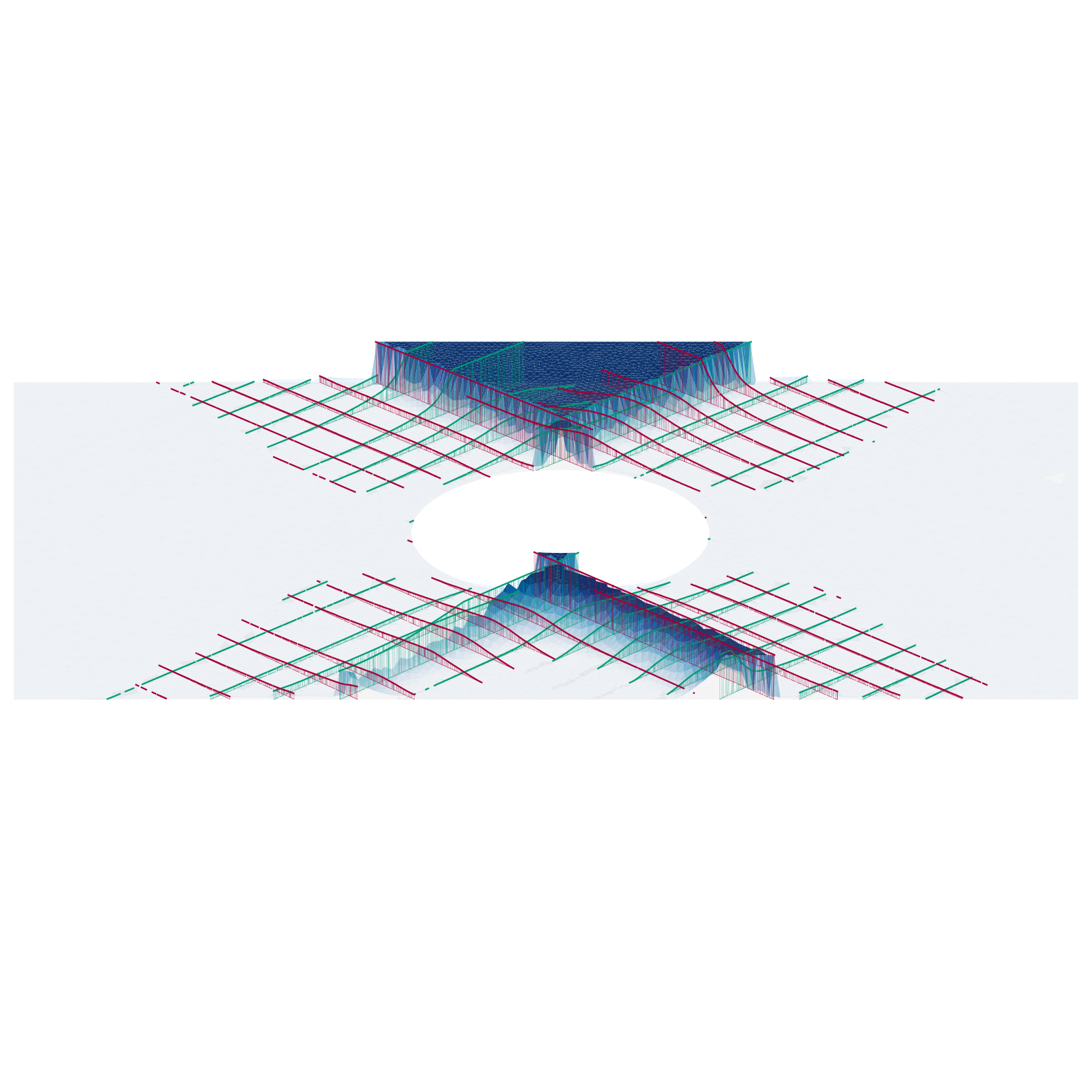}};
    \node[] at (0.6,-5.2){(d)};
    \node at (-4,-7.7) {\includegraphics[clip, width=0.5\columnwidth, trim = 0 3.5cm 0 5cm]{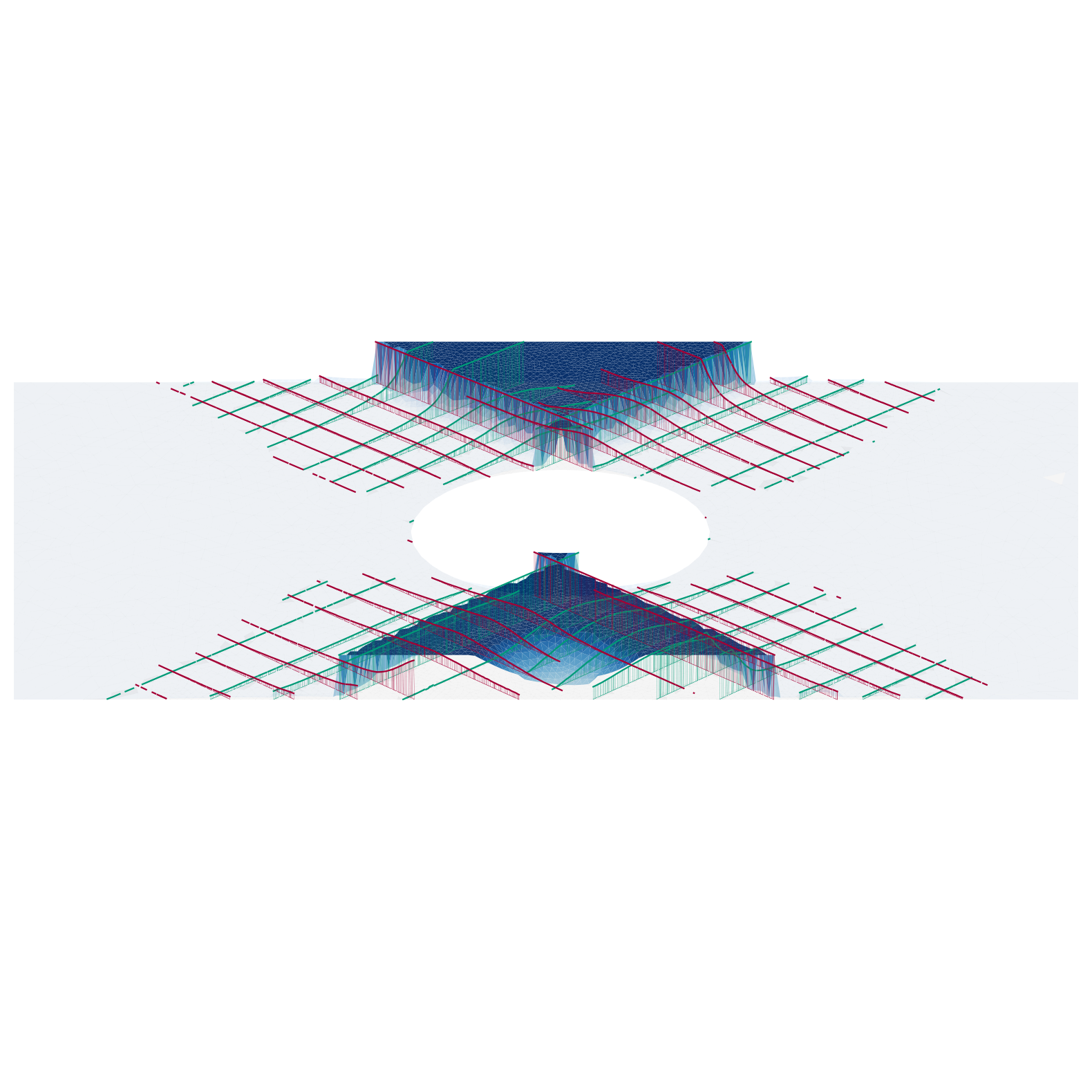}};
    \node[] at (-7.4,-8.2){(e)};
    \node at (4,-7.7) {\includegraphics[clip, width=0.5\columnwidth, trim = 0 3.5cm 0 5cm]{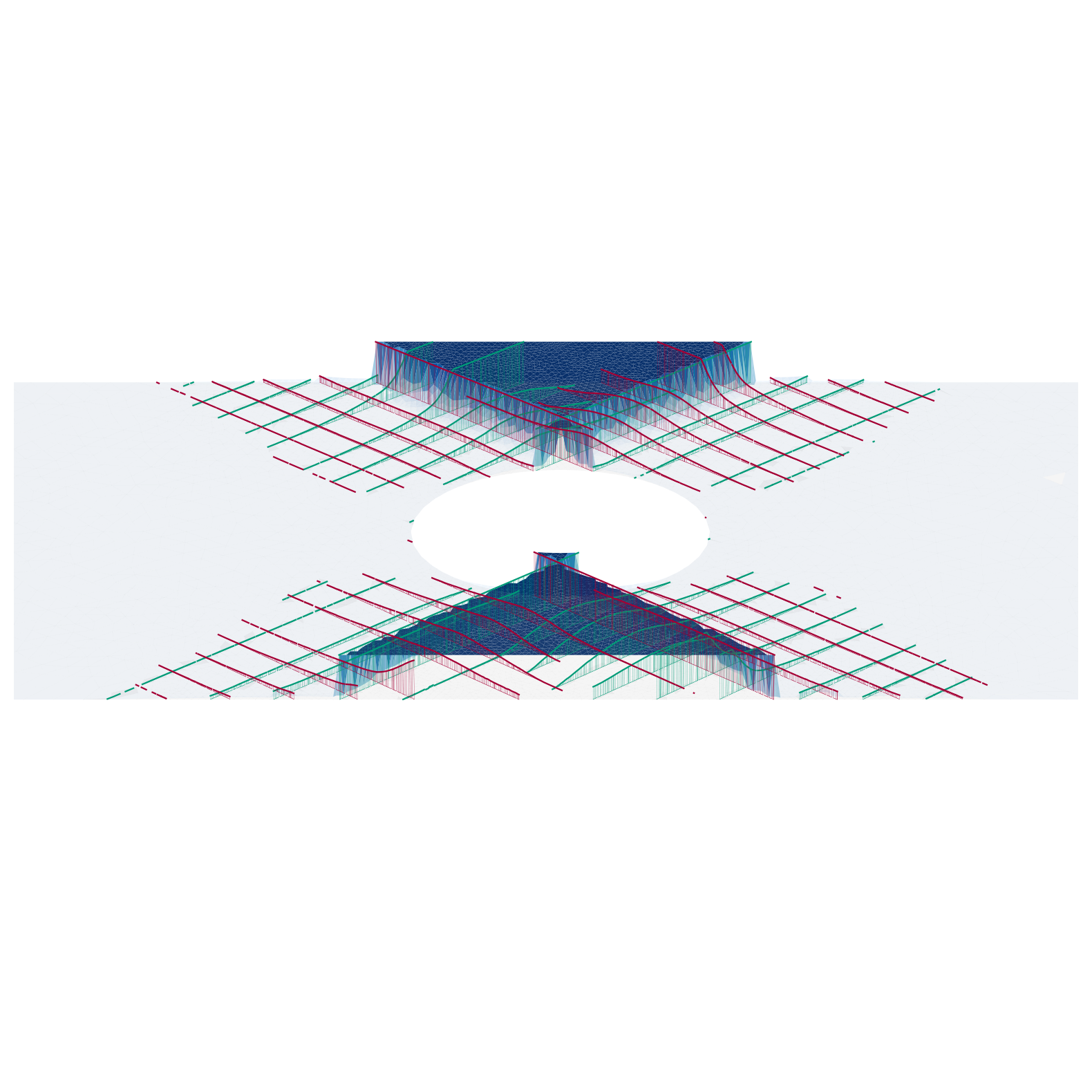}};
    \node[] at (0.6,-8.2){(f)};

    \node (a) at (-2.3,-0.30) {};
    \node (b) at (-2.3,-0.03) {};
    \node[below right=-0.2 and -0.25 of b, font=\fontsize{1pt}{1pt}] (btext) {\tiny$0$};
    \node[below right=-0.8 and -0.25 of a, font=\fontsize{1pt}{1pt}] (btext) {\tiny$1$};
    \node[below left=-0.20 and -0.1 of a] (a1) {};
    \node[below left=-0.20 and -0.1 of b] (b1) {};

    \draw [very thin,-,black] (a.north) to (b.north);
    \draw [very thin,-,black] (a1.north) to (a.north);
    \draw [very thin,-,black] (b1.north) to (b.north);
  \end{tikzpicture}

  \caption{Stiffness reduction as a function of number of cycles $N$ (\emph{top}) and damage evolution at indicated time instances in interface (in \emph{blue}) and XFEM matrix cracks in top ply (in \emph{red}) and bottom ply (in \emph{green})}
  \label{fig:delamination-evol}
\end{figure}

\begin{figure}
  \centering
    \begin{tikzpicture}
      \node at (0,0) {\def\svgwidth{0.5\columnwidth}{\scalebox{1.0}{
        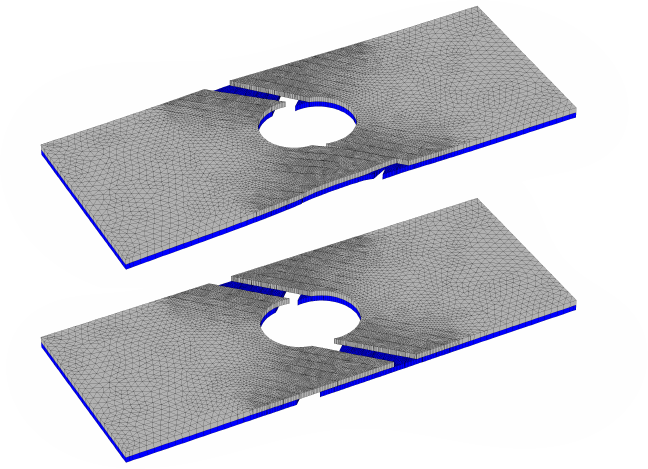}}};
      \node (a) at (4.5,2.2) {$N=1609312$};
      \node (b) at (4.5,0.0) {$N=1759513$};
  \end{tikzpicture}
  \caption{Deformations in the laminate showing a transition from distributed to localized failure}
  \label{fig:deformation}
\end{figure}

\subparagraph*{Global response and damage development}
 Simulations are performed with crack-spacing parameter $l_c = 0.9$. The progressive nature of damage development is depicted in \cref{fig:delamination-evol}, which shows the reduction of \emph{global} stiffness (reaction force over applied displacement at the end) as a function of cycle number $N$. At the indicated time instances, the damage profiles in the interface and the XFEM matrix cracks are shown. From these figures, several stages of damage development can be distinguished. Firstly, a gradual stiffness reduction takes place due to matrix cracking and delamination of the (small) triangular areas near the hole, followed by a rapid damage development (from (a) to (c)) due to combined matrix cracking and major delamination on one side of the hole. After this phase, a third stage of matrix cracking on the opposite side takes place (from (d) to (f)), leading to delamination with sharp stiffness drops until complete failure of the laminate. The deformed mesh is shown in \cref{fig:deformation}, from which it can be observed that a transition from distributed to localized failure is obtained.

\begin{figure}
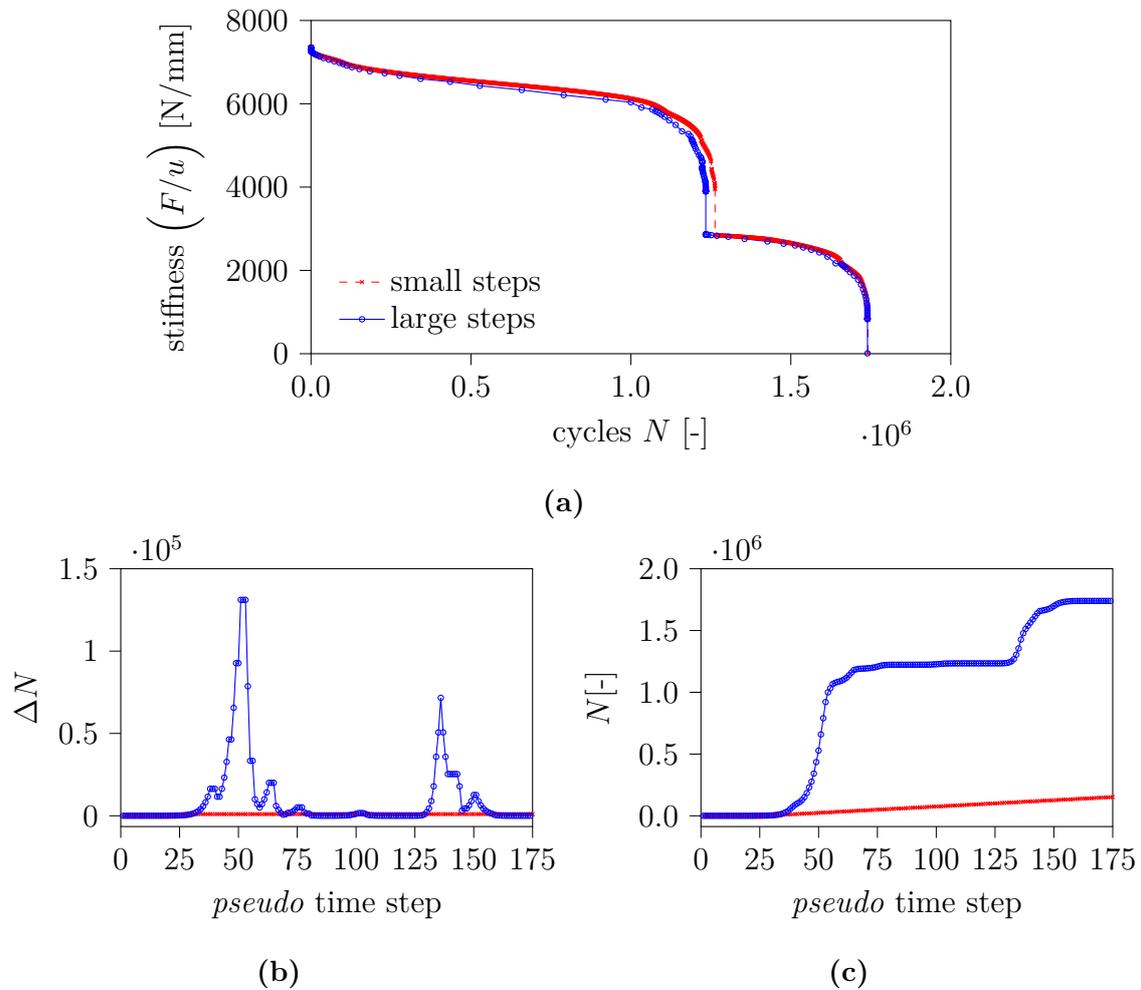

\centering
\begin{subfigure}{1.0\textwidth}
\centering
  \begin{tikzpicture}
  \node[] at (0,0)
    {
      \input{figures/time-dependence.tex}
    };
  \end{tikzpicture}
  \caption{}
  \label{fig:timestep-dependence-stiff}
\end{subfigure}

\begin{subfigure}{0.5\textwidth}
   \begin{tikzpicture}
  \node[] at (-2.5,0)
    {
      \input{figures/cyclejumps.tex}
    };
  \end{tikzpicture}
  \caption{}
  \label{fig:cycle-jumps}
\end{subfigure}%
\begin{subfigure}{0.5\textwidth}
   \begin{tikzpicture}
  \node[] at (2.5,0)
    {
      \input{figures/cycles-vs-timesteps.tex}
    };
  \end{tikzpicture}
  \caption{}
  \label{fig:cycle-jumps}
\end{subfigure}
\caption{Influence of the cycle jump size. \textbf{(a)} stiffness reduction with two different cycle jump parameters corresponding to small and large step sizes. \textbf{(b)} cycle increments per \emph{pseudo} time step. \textbf{(c)} accumulation of fatigue cycles throughout the simulation steps. The simulation with the small step strategy required more than 1500 time steps, whereas with the larger step sizes the simulation finished in 174 time steps} 
\label{fig:timestep-dependence}
\end{figure}

\subparagraph*{Time step dependence}
 The influence of the adaptive cycle jumping scheme on the accuracy of the simulation is illustrated in \cref{fig:timestep-dependence}, where the stiffness reduction is shown in \cref{fig:timestep-dependence-stiff} with two different adaptive stepping parameters, corresponding to small and large cycle increments as depicted in \cref{fig:cycle-jumps}. For the small increment simulation, the maximum cycle jump is $\Delta N^\mathrm{max} = 1000$. It can be observed that the progression of damage and the time to failure is only slightly affected by the large step sizes, thereby validating the choice to use \emph{global} iterations as a measure for determining the cycle increment $\Delta N$ with implicit time integration of the fatigue damage rate function. Moreover, \cref{fig:cycle-jumps} shows that the cycle jumping strategy with consistent tangent and implicit fatigue damage update allows a seamless transition through periods that require small steps and periods that allow for large cycle increments, with minimal influence on the accuracy.
 
\begin{figure}
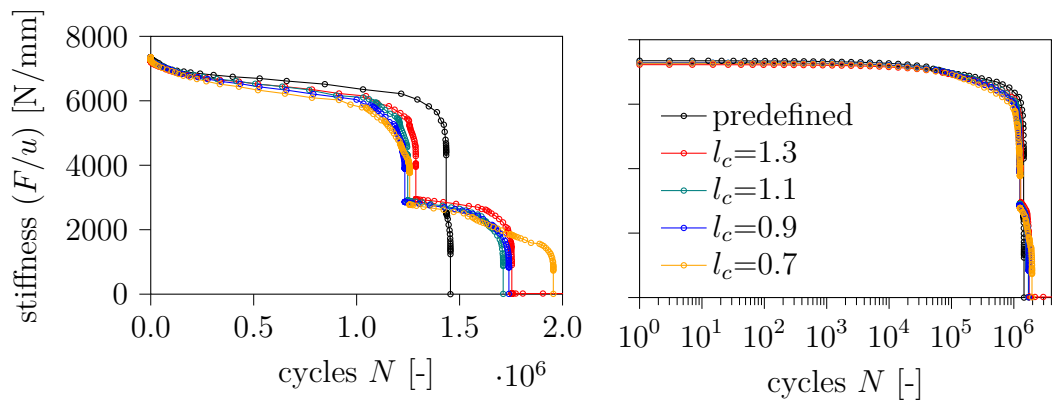

  \centering 
  \begin{tikzpicture}
    \node[] at (-4,0)
    {
      \input{figures/spacing.tex}
    };
    \node[] at (3,-0.28)
    {
      \input{figures/spacing-log.tex}
    };
  \end{tikzpicture}
  \caption{Influence of the spacing parameter $l_\mathrm{c}$ on the simulation results on two different scales: linear on the left and log-scale on the right.} 
  \label{fig:spacing}
\end{figure}

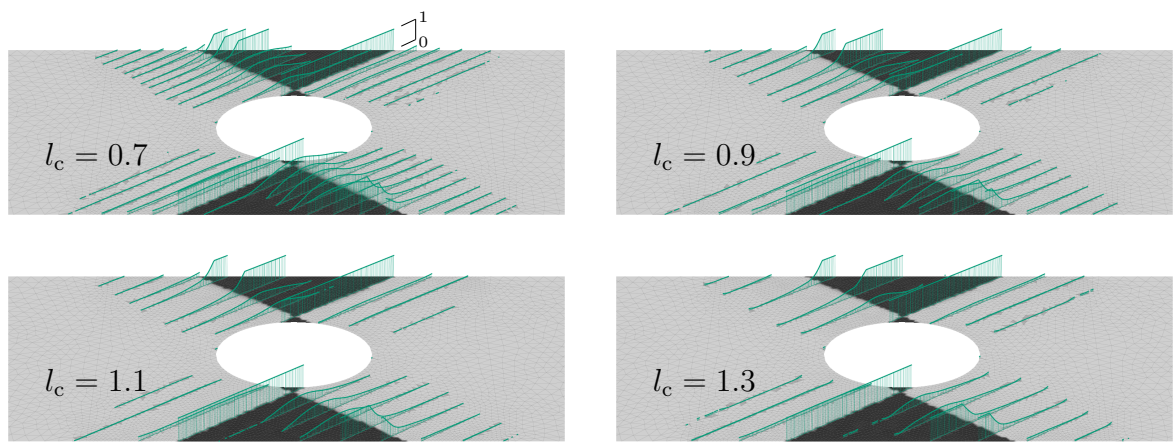
\begin{figure}
  \centering 
  \begin{tikzpicture}
    \node at (-4,-1.7){\includegraphics[clip, width=0.5\columnwidth, trim = 0 3.5cm 0 5cm]{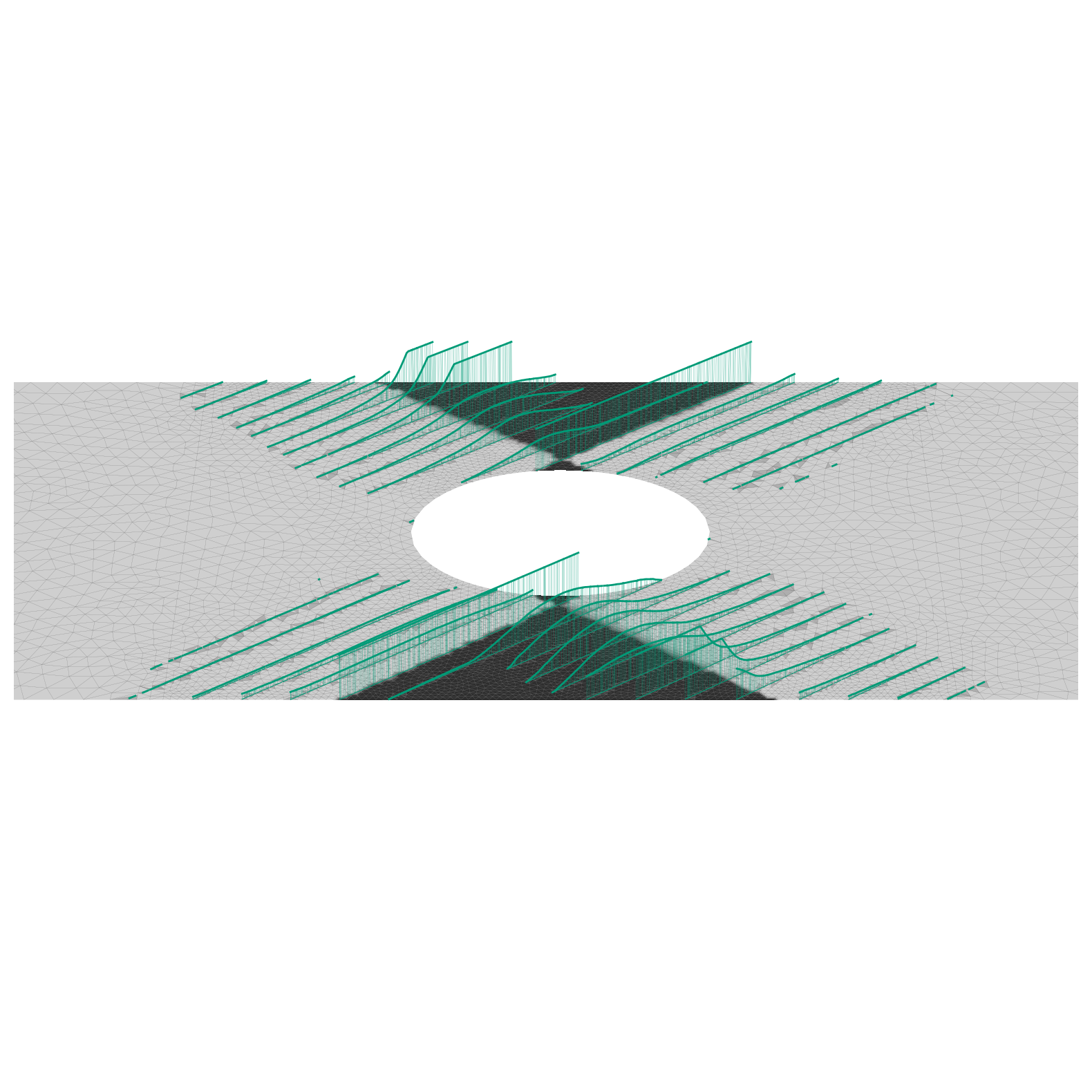}};
    \node at (-6.5,-1.7){$l_\mathrm{c}=0.7$};
    \node at (4,-1.7) {\includegraphics[clip, width=0.5\columnwidth, trim = 0 3.5cm 0 5cm]{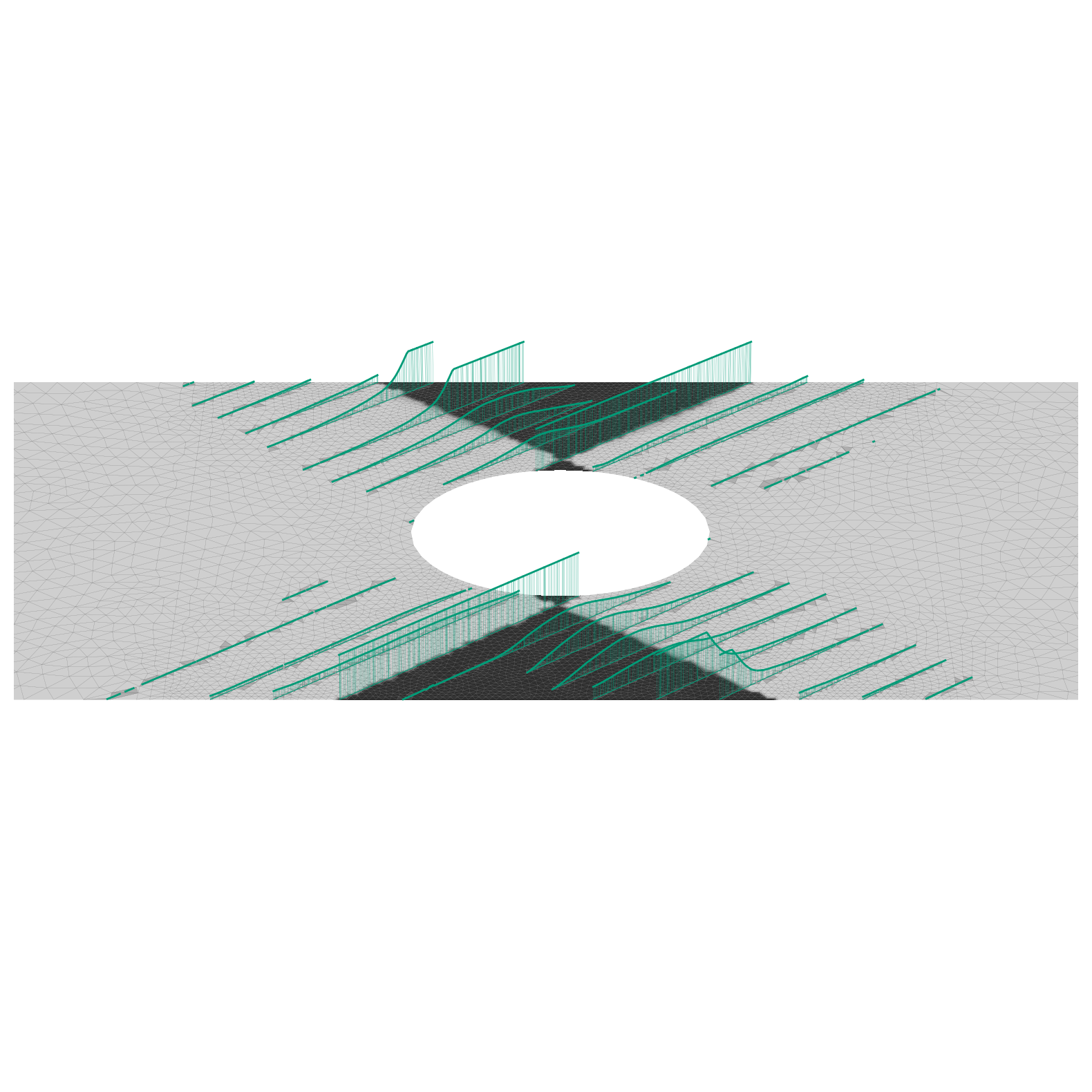}};
    \node at (1.5,-1.7){$l_\mathrm{c}=0.9$};
    \node at (-4,-4.7){\includegraphics[clip, width=0.5\columnwidth, trim = 0 3.5cm 0 5cm]{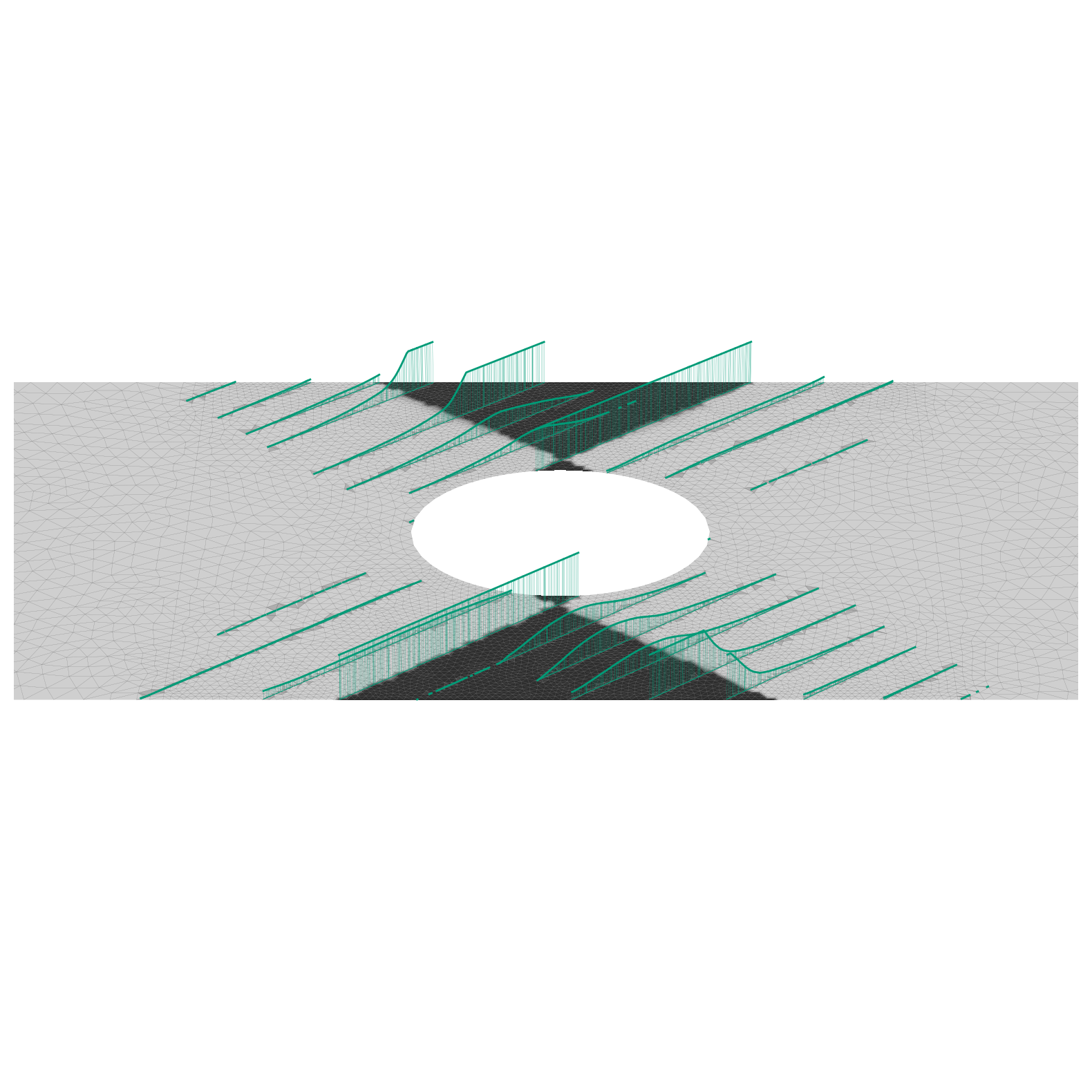}};
    \node at (-6.5,-4.7){$l_\mathrm{c}=1.1$};
    \node at (4,-4.7) {\includegraphics[clip, width=0.5\columnwidth, trim = 0 3.5cm 0 5cm]{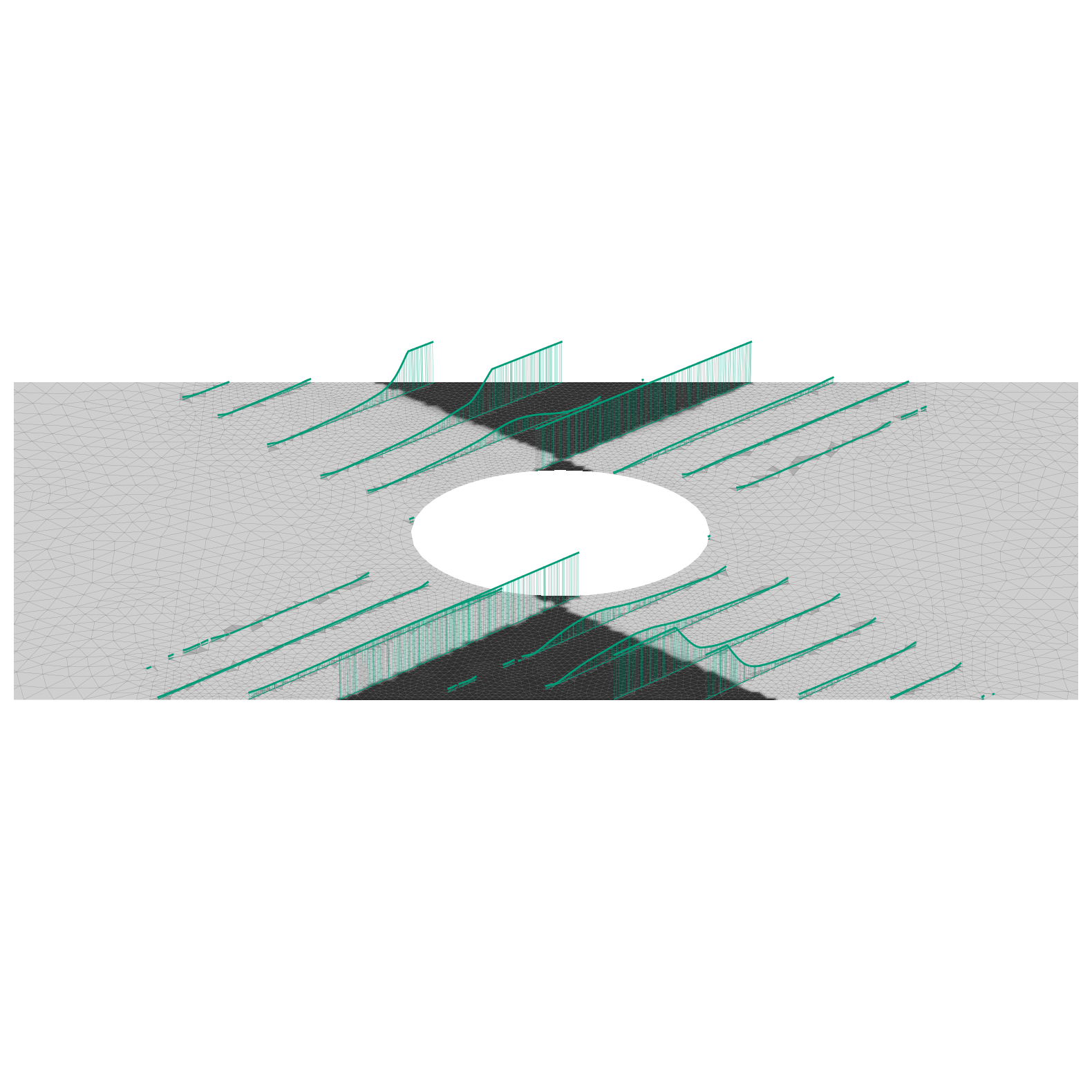}};
    \node at (1.5,-4.7){\textcolor{black}{$l_\mathrm{c}=1.3$}};

    \node (a) at (-2.3,-0.30) {};
    \node (b) at (-2.3,-0.03) {};
    \node[below right=-0.2 and -0.25 of b, font=\fontsize{1pt}{1pt}] (btext) {\tiny$0$};
    \node[below right=-0.8 and -0.25 of a, font=\fontsize{1pt}{1pt}] (btext) {\tiny$1$};
    \node[below left=-0.20 and -0.1 of a] (a1) {};
    \node[below left=-0.20 and -0.1 of b] (b1) {};

    \draw [very thin,-,black] (a.north) to (b.north);
    \draw [very thin,-,black] (a1.north) to (a.north);
    \draw [very thin,-,black] (b1.north) to (b.north);
  \end{tikzpicture}
  \caption{Final damage $\Dam$ in XFEM cracks in bottom ply (shown in \emph{green}) for different crack spacing parameter values $l_\mathrm{c}$. The delaminated area in the interface is shown in \emph{dark gray}} 
  \label{fig:spacing-xfem-damage}
\end{figure}

\subparagraph*{Crack-spacing parameter dependence}
The exercise is repeated for several crack-spacing parameter values $l_c(\mathrm{mm}) \in \{0.7,0.9,1.1,1.3\}$. Moreover, a case with two cracks per ply at predefined locations is included. The cracks are inserted in each ply in the direction of the fibers and tangential to the hole. The \emph{global} stiffness reduction is shown in \cref{fig:spacing}. An interesting observation is that allowing more than two cracks per ply has a significant positive effect on the fatigue life of the considered laminate. This can be explained by the fact that matrix cracking results in a reduction of stress concentrations in the interface and thus less fatigue damage accumulation. Another observation is that the response varies slightly with different crack-spacing parameter values. However, the discrepancy is not monotonic, in the sense that reducing the spacing does not always lead to an increase in fatigue life. In fact, the difference can be explained by analysing \cref{fig:spacing-xfem-damage}, which depicts the damage in the XFEM matrix cracks in the bottom ply for each spacing parameter value $l_\mathrm{c}$. It can be observed that the crack density and patterns are different, which explains a different progression of stiffness reduction as observed in \cref{fig:spacing}. 
However, when the response is shown on a log-scale, as is common practice with fatigue analysis, the difference vanishes. 

\begin{figure}
  \centering 
  \begin{tikzpicture}
    \node at (-4,0){\includegraphics[clip, width=0.6\textwidth]{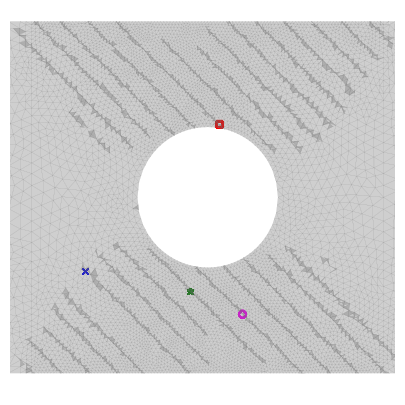}};
    \node at (4,-0.5) {\input{figures/ipoints.tex}};
  \end{tikzpicture}
  \caption{Traction-opening histories (\emph{right}) of four XFEM integration points at different crack locations in the ply (\emph{left})} 
  \label{fig:ipoints}
\end{figure}

\Cref{fig:ipoints} shows the response in four integration points for the case of a crack spacing parameter $l_c=\SI{0.9}{mm}$. As it can be observed, two of the selected points completely separate, while the other integration points unload. This indicates that not necessarily all XFEM cracks that have been inserted accumulate damage until complete material point failure, thereby allowing for simulating the transition from distributed damage to localized failure.
Another interesting fact is that the traction-separation response is not parallel to the static softening line, as is the case with the simulated DCB case in the previous section (cf. \cref{fig:tvsu-implicit}), indicating the accurate time integration of the damage evolution equation.

\section{Conclusion}

A numerical framework for simulating progressive fatigue failure has been presented in this paper. The recently proposed fatigue cohesive zone model by \citeauthor{Davila2020}, which covers initiation and propagation, has been improved with an implicit time integration scheme and consistent linearization of both the underlying quasi-static and the fatigue cohesive relation. Furthermore, the fatigue cohesive zone model has been combined with XFEM for modeling mesh-independent transverse matrix cracks in full-laminate analyses.

It has been shown with numerical examples that the improved damage update results in more accurate and efficient analyses.
The capabilities of the numerical framework have been demonstrated with the simulation of an open-hole [$\pm45$]-laminate under fatigue loading. The numerical model can accurately simulate the interaction of transverse matrix cracking and delamination. A slight sensitivity to the numerical crack-spacing parameter has been observed, although within acceptable range for predicting fatigue life.

\section*{Acknowledgement}
This research was carried out as part of the project ENLIGHTEN (project number N21010h) in the framework of the Partnership Program of the Materials innovation institute M2i (www.m2i.nl) and the Netherlands Organization for Scientific Research (www.nwo.nl).

\printbibliography[heading=bibintoc]

\end{document}